\documentclass[conference]{IEEEtran}
\IEEEoverridecommandlockouts

\usepackage{cite}
\usepackage{amsmath,amssymb,color,graphicx,caption,subcaption,comment,tikz,url,latexsym} 
\usepackage{algorithmic}
\usepackage{graphicx}
\usepackage{textcomp}
\usepackage{pgfplots}
\usepackage{xcolor}
\usepackage{url}
\usepackage{multirow}
\usepackage{mathtools}
\usepackage[left=0.62in,right=0.62in,top=0.6in,bottom=0.6in]{geometry}

\usepgfplotslibrary{units}
\usetikzlibrary{spy,backgrounds}
\usepackage{pgfplotstable}

\def\BibTeX{{\rm B\kern-.05em{\sc i\kern-.025em b}\kern-.08em
		T\kern-.1667em\lower.7ex\hbox{E}\kern-.125emX}}

\usepackage{cite}
\usepackage{amsmath,amssymb,color,graphicx,caption,subcaption,comment,tikz,url,latexsym} 
\usepackage{graphicx}
\usepackage{textcomp}
\usepackage{pgfplots}
\usepackage{xcolor}
\usepackage{url}
\usepackage{multirow}
\usepackage{bbm}
\usepackage{dsfont}
\usepackage{mathtools}
\usepgfplotslibrary{units}
\usetikzlibrary{spy,backgrounds}
\usepackage{pgfplotstable}

\def\BibTeX{{\rm B\kern-.05em{\sc i\kern-.025em b}\kern-.08em
		T\kern-.1667em\lower.7ex\hbox{E}\kern-.125emX}}

\newtheorem{theorem}{Theorem} 

\newtheorem{definition}{Definition}

\newtheorem{corollary}[theorem]{Corollary}
\newtheorem{remark}{Remark}

\usepackage{bbm}

\begin{document}
	\interdisplaylinepenalty=0
	\title{Distributed Detection under Stringent Resource Constraints}
	\allowdisplaybreaks[4]
	\sloppy
	
    \author{
    \IEEEauthorblockN{Abdelaziz Bounhar$^{\star}$, Mireille Sarkiss$^{\S}$, and Mich\`ele Wigger$^{\dagger}$\vspace{1mm}}
    
        \IEEEauthorblockA{$\star$MBZUAI, Paris, France. Email: abdelaziz.bounhar@mbzuai.ac.ae} 
                \IEEEauthorblockA{$^{\S}$SAMOVAR, T\'{e}l\'{e}com SudParis, Institut Polytechnique de Paris, 91120 Palaiseau, France\\ Email: mireille.sarkiss@telecom-sudparis.eu}
       \IEEEauthorblockA{$^{\dagger}$LTCI, T\'{e}l\'{e}com Paris, Institut Polytechnique de Paris, 91120 Palaiseau, France, Email: 
michele.wigger@telecom-paris.fr}
	
}

\author{Abdelaziz Bounhar, Mireille Sarkiss, Mich\`{e}le Wigger
\thanks{This work was presented in part at the 2025 IEEE Information Theory Workshop. A. Bounhar was with T\'{e}l\'{e}com Paris and is now with MBZUAI, Paris, France. Email: abdelaziz.bounhar@mbzuai.ac.ae. Mireille Sarkiss is with SAMOVAR, T\'{e}l\'{e}com SudParis, Institut Polytechnique de Paris, 91120 Palaiseau, France. Email: mireille.sarkiss@telecom-sudparis.eu. Michèle Wigger is with LTCI, T\'{e}l\'{e}com Paris, Institut Polytechnique de Paris, 91120 Palaiseau, France, Email: 
michele.wigger@telecom-paris.fr.}
}
	\maketitle

\begin{abstract}
This paper identifies the Stein-exponent of distributed detection when the sensor communicates to the decision center over a discrete memoryless channel (DMC) subject to one of three stringent communication constraints: 1) The number of channel uses of the DMC grows sublinearly in the number of source observations $n$;  2) The number of channel uses is $n$ but a block-input cost constraint $C_n$ is imposed  almost surely and $C_n$ grows sublinearly in $n$; 3) The block-input constraint is imposed only on expectation.  We identify a dichotomy in the Stein-exponent of all these setups depending on the structure of the DMC's transition law. Under  any of these three constraints, when the DMC is partially-connected (i.e., some outputs cannot be induced by certain inputs) the Stein-exponent matches the exponent identified by Han and Kobayashi and by Shalaby and Papamarcou for the scenario where communication is of  zero-rate but over a noiseless link. We prove our   result   by adapting Han's zero-rate coding strategy  to partially-connected DMCs. 
  In contrast, for fully-connected DMCs, in our scenarios 1) and 2) the Stein-exponent collapses to that of a local test at the decision center, rendering the remote sensor  and  communication useless. 
   In scenario 3), the sensor remains beneficial even for fully-connected DMCs, however also collapses compared to the case of a partially-connected DMC. Moreover,  the Stein-exponent is larger when the expectation constraint is imposed only under the null hypothesis compared to when it is imposed under both hypotheses. To prove these results, we propose both new coding strategies and new converse proofs. 

\end{abstract}
\section{Introduction}
Consider a distributed binary hypothesis testing problem where a sensor and a decision center observe correlated sources and the  sensor can  transmit information to the decision center over a channel.  The joint distribution of the observations at the sensor and the decision center depends on a binary hypothesis $\mathcal{H}\in\{0,1\}$ that should be guessed by the decision center based on its own observed samples and the outputs of the channel.   The goal is to reduce the error probabilities under the two hypotheses. In particular,  we wish to  keep the  type-I error probability (deciding $\hat{\mathcal{H}}=1$ if $\mathcal{H}=0$) below a given threshold $\epsilon\in[0,1)$ while driving the type-II error probability (deciding $\hat{\mathcal{H}}=0$ if $\mathcal{H}=1$) to 0 exponentially fast in the number of observation samples $n$. This type of asymmetric constraint is particularly relevant in alert systems, where keeping the false alarm rate below a certain threshold is sufficient, but minimizing the missed detection probability is critical. The largest exponential decay of the type-II error probability in such an asymmetric setup is known as the Stein-exponent. 

This canonical distributed hypothesis testing problem has been studied in various previous works under different assumptions on the communication channel.  An important line of work started in \cite{Ahlswede} and continued in \cite{Han, SHA, Wagner, Kochman-MAC, WeinbergerKochman, KochmanWang, SadafMicheleLigong}; it studied the Stein-exponent in a communication scenario where the sensor can send $R n$ bits to the decision center over an error-free link. Achievability results were reported, but the exact exponent is known only for special classes of source distributions. Extensions to network scenarios with multiple sensors and/or decision centers have been reported since, see \cite{Wagner, salehkalaibar2020hypothesisv1, Michele3}
, as well as extensions to include security constraints \cite{Mhanna, 8125176, 8664261, TACI_HT, Faour, CovertDHT}, or to allow for variable-length coding \cite{HWS20, JSAIT}.

A somehow separate  line of work \cite{Han_Kobayashi, Shalaby, PierreMichele, Watanabe_DHT}  derived the Stein-exponent of binary distributed hypothesis testing in a scenario where the number of communication bits from the sensor to the decision center is only sublinear in the number of observations  $n$ and not linear in $n$ as assumed in the works above. Such a scenario is commonly referred to as zero-rate communication.  Under this assumption, \cite{Han_Kobayashi, Shalaby} characterized  the Stein-exponent for a large class of distributions. Specifically, Han and Kobayashi \cite{Han_Kobayashi} focused on the scenario where the sensor can only send a single bit, and showed that in this  case  the optimal strategy  is that the  decision center decides on $\hat{\mathcal{H}}=0$ if, and only if,  its own local observation and the sensor's observations are each \emph{marginal}-typical according to the distributions under hypothesis $\mathcal{H}=0$. Implementing this strategy requires only that the sensor sends the binary outcome of its local typicality test to the decision center.
This simple 1-bit communication strategy and the corresponding strategy at the decision center were proved to be  optimal and achieve the Stein-exponent  by Shalaby and Papamarcou \cite{Shalaby} for all setups where the sensor can send   a sublinear number of noise-free bits to the decision center. These zero-rate results have recently also been extended to the quantum case \cite{Sreekumar}.


While above described results considered error-free communication channels (except for \cite{CovertDHT}), in this work, we assume that communication from the sensor to the decision center takes place over a discrete memoryless channel (DMC). Distributed hypothesis testing over a DMC  has already been considered in  \cite{Deniz_DHT, Michele_noisy_and_MAC}  under the assumption that the DMC can be used $\kappa n$ times, i.e., the number of channel uses is linear in the observation length $n$.  The authors in \cite{Deniz_DHT}  determined the Stein-exponent  for the class of source distributions termed ``testing against conditional independence" and showed that it coincides with the exponent of a noise-free link of same capacity. The fact that the DMC is noisy thus does not decrease the Stein-exponent in these setups. For other source distributions however this is not the case and the noise in the DMC  decreases the Stein-exponent  and mechanisms such as ``unequal error protection" need to be employed to partially mitigate the noise   \cite{Deniz_DHT, Michele_noisy_and_MAC}. Similar results were also reported for the MAC \cite{Michele_noisy_and_MAC} and the BC \cite{Sadaf_BC}.

In this work, we study the influence of the noise in a DMC on the zero-rate-results by Shalaby and Papamarcou  \cite{Shalaby}. 
Specifically, we 
show a dichotomy of the Stein-exponent depending on whether the transition diagramme of the DMC is fully-connected (i.e., each input generates each output with positive probability) or not. For fully-connected DMCs,  the Stein-exponent completely collapses to the  exponent with only local observations at the decision center without any communication  from the sensor. The  sensor thus becomes useless in this case and  a local test at the decision center attains the Stein-exponent. In contrast, when the DMC is only partially-connected, it is possible to attain the same  Stein-exponent  as when communication is over a noiseless link.  
As before, the exponent is achieved by having the decision center declare $\hat{\mathcal{H}}=0$ only when  the  marginal typicality tests for the distribution under $\mathcal{H}=0$ are satisfied both at the sensor and at the decision center. Over a noisy but partially-connected DMC, this decision strategy can be implemented by having the sensor send $k$ times symbol $x_0$ if its own marginal typicality check is successful and $k$ times a different symbol $x_1$ otherwise, where $x_0$ and $x_1$ are chosen so that a certain output  $y^*$ is never reached  by input $x_1$. The decision center then only declares $\hat{\mathcal{H}}=0$ if it observes at least one output $y^*$ and its local typicality check is satisfied.  This way,  the type-II error is not increased compared to the noiseless link scenario. Moreover, for increasing  $k\to \infty$ the type-I error vanishes as for noise-free communication.  For finite number of channel uses $k$ the strategy remains applicable if the tolerated type-I error probability is sufficiently large. 

We further prove the same results on the Stein-exponent in a related setup where the DMC can be used $n$ times but a block-input power constraint $C_n$ is imposed that grows only sublinearly in $n$. (General cost constraints are allowed as long as there exists exactly one symbol with zero costs.) 

In our third setup, we relax the cost constraint to hold only \emph{on expectation} and determine the exact Stein-exponent for this setup. For partially-connected DMCs, the Stein-exponent and the optimal coding scheme are as before. For fully-connected DMCs, we propose a coding scheme that improves on the local Stein-exponent at the decision center, and the sensor is thus not useless. In fact, under the expected cost constraint, the sensor can still  send codewords with large costs, albeit with very small probabilities. We can therefore employ the previously described scheme with the marginal typicality checks  where  the sensor sends the  zero-cost symbol  {over the entire blocklength} $n$ in case its local typicality test succeeds, and otherwise  it  sends any other symbol $x_1$ \emph{over the entire blocklength}. Since the typicality test fails with exponentially vanishing probability, this strategy will not violate the expected cost constraint. It however achieves a strictly better exponent than the local test at the decision center. By proposing two converse proofs based on carefully chosen change-of-measure arguments, we show that the described coding and decision strategy achieves indeed the optimal Stein-exponent. This Stein-exponent not only depends on the source distributions but also on the  largest error exponent that can be attained over the DMC when transmitting a binary symbol. 

In this last setup, we distinguish between the two scenarios where the expected cost constraint is imposed under both hypotheses or only under hypothesis $\mathcal{H}=0$. The single-hypothesis constraint is motivated by highly critical alarm systems where in case of alarm  situation the consumed power is of no interest. 
  It turns out that the Stein-exponent is sensitive to this distinction and deteriorates when the cost constraint is imposed under both hypotheses.  In fact, when the expected cost constraint is also imposed under $\mathcal{H}=1$, then the sensor also needs to send  the zero-cost symbol when its observation is marginal-typical according to the distribution under $\mathcal{H}=1$, and otherwise it suffices to send it when it is typical according to the distribution under $\mathcal{H}=0$.

To summarize, in this work, we reexamine the results on the Stein-exponent under zero-rate communication in \cite{Shalaby},  now with communication over a DMC. We establish the Stein-exponents for three communication scnearios: 1) The number of channel uses of the DMC grows sublinearly in the number of source observations $n$.  2) The number of channel uses is $n$ but a block-input cost constraint $C_n$ is imposed  almost surely and $C_n$ is sublinear in $n$; 3) The stringent block-input constraint  $C_n$ is imposed only on expectation.  Our results reveal a dichotomy of the Stein-exponent in these setups depending on whether the DMC is fully- or partially-connected. When the DMC is partially-connected,  it is possible to achieve the same Stein-exponent as reported in   \cite{Shalaby} for noiseless channels.  In contrast, when the channel is fully-connected, the optimal Stein-exponent collapses. While in cases 1) and 2) the exponent collapses  to the local exponent at the decision center  and  the sensor is rendered useless, in case 3) the collapse is only partial and the sensor still helps improving the Stein-exponent. In this case 3), the Stein-exponent moreover depends on the exact channel law and on whether the expected cost constraint is imposed under both hypotheses or only under the null hypothesis.

\textit{Notation:}
We mostly follow standard notation. In particular, random variables are denoted by upper case letters (e.g., $X$), while their realizations are denoted by lowercase (e.g. $x$). We  abbreviate  $(x_1,\ldots, x_n)$ by  $x^n$.  We use the function $w_{\textnormal{H}}(\cdot)$ 
to indicate the Hamming weight. 
We abbreviate \emph{independent and identically distributed} as \emph{i.i.d.} and \emph{probability mass function} as \emph{pmf}. 
Also, we denote by $\pi_{u^nv^n}$ the joint type of the sequences $(u^n, v^n)$:
\begin{equation}
\pi_{u^nv^n}(a,b)\triangleq\frac{|\{i\colon (u_i,v_i)=(a,b)\}|}{n}, \quad (a,b)\in  \mathcal{U}\times \mathcal{V},
\end{equation}
and $\pi_{u^n}(a)$ the corresponding marginal type of the $u^n$ sequence.
We use $\mathcal{T}_{\mu}^{(n)}(P_{U})$ to denote the jointly strongly-typical set as in \cite[Definition 2.9]{Csiszarbook}, i.e., the set of all sequences $u^n$ with type $\pi_{u^n}$ satisfying $\pi_{u^n}(a)=0$ if $P_U(a)=0$ and   $|\pi_{u^n}(a)-P_U(a)|\leq \mu$ otherwise..
We also use Landau notation $o(1)$ to indicate any function that tends to 0 for blocklengths $n\to \infty$.

\section{Distributed Detection with a Sublinear Number of Channel Uses}

\subsection{Problem Setup} \label{sec:HT}

\begin{figure}[t!]
\begin{center}
\includegraphics[scale=0.3]{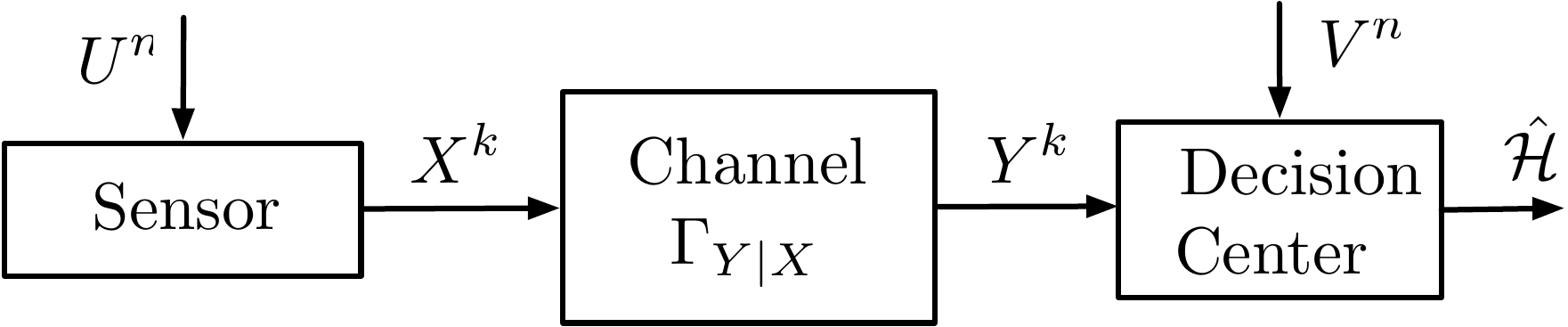}
\end{center}
\caption{Distributed Hypothesis testing with a sublinear number of channel uses.}
\label{fig:DHT_sublinear}
\vspace{-1mm}
\end{figure}
Consider the distributed hypothesis testing problem  in Figure~\ref{fig:DHT_sublinear} where  for a given blocklength $n$, a sensor observes a sequence $U^n$ and communicates to a decision center with local observations $V^n$.  
The  
 distribution of the observations $(U^n,V^n)$ depends  on a binary hypothesis $\mathcal{H}\in\{0,1\}$:
\begin{subequations}
	\begin{IEEEeqnarray}{rCl}
		& &\textnormal{if } \mathcal{H} = 0 \colon \quad (U^n,V^n ) \textnormal{ i.i.d. } \sim \, P_{UV}; \\
		& &\textnormal{if } \mathcal{H} = 1\colon  \quad (U^n,V^n)  \textnormal{ i.i.d. } \sim\, Q_{UV},
	\end{IEEEeqnarray} 
\end{subequations}
for  given pmfs $P_{UV}$ and $Q_{UV}$ over  the product alphabet $\mathcal{U}\times \mathcal{V}$, where we assume that $Q_{UV}(u,v)>0$ for all $(u, v)\in\mathcal{U}\times \mathcal{V}$. Let $P_{U}$ and $P_{V}$ denote the marginal pmfs of $P_{UV}$. Similarly to the results in \cite{Shalaby}, we assume that the support of $P_{UV}$ is included in the support of $Q_{UV}$.

Communication from   sensor  to  decision center takes place over $k(n)$ uses of a discrete memoryless channel DMC 
with finite input and output alphabets $\mathcal{X}$ and $\mathcal{Y}$ and  transition law $\Gamma_{Y|X}$.  The number of channel uses grows sublinearly in $n$: 
\begin{subequations}\label{eq:sublinear}
\begin{IEEEeqnarray}{rCl}
	\lim_{n\to \infty}k(n)&= &\infty \\
	\lim_{n\to \infty}\frac{k(n)}{n}&= &0. 
\end{IEEEeqnarray} 	
\end{subequations}
For ease of notation we will also write $k$ instead of $k(n)$.

The encoder and decoder are thus formalized by two functions $f^{(n)}$ and $g^{(n)}$ on appropriate domains, where $f^{(n)}$ describes how observations $U^n$ are mapped to channel inputs $X^k$:
\begin{equation}
X^k=f^{(n)}(U^n) \in \mathcal{X}^k,
\end{equation}
and $g^{(n)}$ describes how channel outputs $Y^k$ and observations $V^n$ are used to generate the decision $\hat{\mathcal{H}}$:
\begin{equation}\label{eq:guess}
\hat{\mathcal{H}} =	g^{(n)}(V^n,Y^k) \in\{0,1\}.
\end{equation}

The goal is to design encoding and decision functions such
that the type-I (false alarm) error probability 
\begin{equation}
\alpha_n \triangleq \Pr \left[ \hat{\mathcal{H}}=1 |\mathcal{H}=0 \right] 
\end{equation} 
stays below given threshold $\epsilon > 0$ and the type-II (miss-detection)
error probability
\begin{equation}
\beta_n \triangleq  \Pr \left[ \hat{\mathcal{H}}=0 |\mathcal{H}=1 \right]
\end{equation}
decays to 0 with largest possible exponential decay.

 \begin{definition}\label{def:ach2} Given  $\epsilon\in[0,1)$, a miss-detection error exponent $\theta>0$ is called $\epsilon$-achievable  if there exists a  sequence of encoding and decision functions   $\{(f^{(n)},g^{(n)})\}_{n=1}^\infty$  satisfying:
\begin{subequations}\label{eq:criteria2}
	\begin{IEEEeqnarray}{rCl}
	\varlimsup_{n\to \infty}  \alpha_n &\leq&  \epsilon \label{eq:P12}\\
\varliminf_{n\to \infty}- \frac{1}{n} \log \beta_n&\geq&  \theta \label{eq:det_exp2}. 	\end{IEEEeqnarray} 
\end{subequations}
The supremum over all $\epsilon$-achievable  miss-detection error exponents $\theta$ is denoted $\theta_{\textnormal{sublin}}^\star(\epsilon)$ and called the \emph{Stein-exponent}.
 \end{definition}

\subsection{Results}\label{theorem1}
The following theorem determines the  Stein-exponent $\theta_{\textnormal{sublin}}^\star(\epsilon)$, which depends on the source distributions $P_{UV}$ and $Q_{UV}$ as well as on the DMC transition law $\Gamma_{Y|X}$, however not on $\epsilon\in[0,1)$. In particular, the theorem illustrates a dichotomy  of the Stein-exponent with respect to the transition law $\Gamma_{Y|X}$, depending on whether the transition diagramme of the  DMC  is fully-connected (i.e., each input induces each output with positive probability) or only partially-connected (some inputs do not induce some of the outputs.) For partially-connected DMCs, the Stein-exponent coincides with the Stein-exponent when communication takes place over a noise-free channel that can be used $k(n)$ times. For fully-connected  DMCs, the Stein-exponent collapses to the exponent of the local test at the decision center and is thus  obtained by a simple test at the decision center without any communication. In this sense,  partially-connected DMCs are equally-good for distributed detection as noiseless links in the regime where the number of channel uses $k(n)$ is sublinear in $n$, while all fully-connected DMCs are completely useless.  

	\begin{theorem}\label{thm_sublinear}
	Fix $\epsilon \in [0,1)$.
\begin{enumerate}
\item If the DMC is such that there exist two inputs $x_0, x_1\in\mathcal{X}$ and an output $y^*\in \mathcal{Y}$ satisfying the two conditions:
\begin{subequations}\label{eq:channel_cond0}
\begin{IEEEeqnarray}{rCl}
\Gamma_{Y|X}(y^*|x_0) &> &0 \\
\Gamma_{Y|X}(y^*|x_1) &= &0,
\end{IEEEeqnarray}
\end{subequations}
 the largest miss-detection error probability is given by:
\begin{IEEEeqnarray}{rCl}\label{eq:Ia}
\theta_{\textnormal{sublin}}^\star(\epsilon) &=&  \min_{\substack{\pi_{UV} \colon \\
\pi_U=P_U\\
\pi_V=P_V}} D( \pi_{UV} \| Q_{UV}).
\end{IEEEeqnarray}
\item Otherwise, it is given by
\begin{IEEEeqnarray}{rCl}\label{eq:Ib}
\theta_{\textnormal{sublin}}^\star(\epsilon) &=&   D( P_{V} \| Q_{V}).
\end{IEEEeqnarray}
\end{enumerate}	
\end{theorem}	
\begin{IEEEproof}Exponent \eqref{eq:Ib}  is simply achieved with a local test  at the decision center, without any communication. The converse  to \eqref{eq:Ib} is proved in Appendix~\ref{app2}. 

The converse for \eqref{eq:Ia} follows from the result in \cite{Shalaby}, which proves that the largest exponent over a noiseless link cannot exceed the exponent on the right-hand side of \eqref{eq:Ia}. 

To prove achievability of \eqref{eq:Ia}, consider the following scheme. Fix a small number $\mu >0$ and let $x_0,x_1,y^*$ be as in the theorem. 

\noindent\underline{Sensor:} If $U^n \in \mathcal{T}^{(n)}_{\mu}(P_{U})$,  send $X^k=x_0^k$. Otherwise, send $X^k=x_1^k$.

\noindent\underline{Decision Center:} If at least one of the channel outputs is $y^*$ and if  $V^n \in \mathcal{T}^{(n)}_{\mu}(P_{V})$, then declare $\hat{\mathcal{H}}=0$. Otherwise, declare $\hat{\mathcal{H}}=1$. 

The analysis of this scheme in Appendix~\ref{app:1} concludes the achievability proof. \end{IEEEproof}
	\medskip

\begin{remark}[Finite Values of $k$] 
The above theorem holds under the assumption that $k\to \infty$. Most of the results however remain valid also for fixed and finite $k\geq 1$. Specifically, the theorem remains valid for all DMCs violating Condition \eqref{eq:channel_cond0} for all triples $(x_0,x_1, y^*)$. For DMCs satisfying \eqref{eq:channel_cond0} for at least one triple $(x_0,x_1, y^*)$, the converse proof trivially remains valid. By inspecting the proof in Appendix~\ref{app:1}, 
we see that achievability continues to hold for all allowed type-I error probabilities $\epsilon \geq  (1- \Gamma_{Y|X}(y^*|x_0))^k$.
	\end{remark}
\begin{remark} 
Result \eqref{eq:Ib} holds also when the support of $P_{UV}$ is not included in the support of $Q_{UV}$.
\end{remark}
	
\begin{corollary}[When communication never helps]
Consider pmfs $P_{UV}$ and $Q_{UV}$ satisfying 
\begin{equation} \label{eq:cond}
\sum_{v} P_V(v) Q_{U|V}(u|v) = P_U(u), \quad \forall u\in \mathcal{U}.
\end{equation}
Then, for any $\epsilon \in [0,1)$, and irrespective of the DMC $\Gamma_{Y|X}$: 
\begin{IEEEeqnarray}{rCl}\label{eq:exact}
\theta_{\textnormal{sublin}}^\star(\epsilon)  &=& D( P_{V} \| Q_{V}).
	\end{IEEEeqnarray}
In particular, for testing against independence ($Q_{UV}=P_{U}\cdot P_{V}$) the Stein-exponent is zero. 

Thus, under Condition \eqref{eq:cond} the  dichotomy observed in Theorem~\ref{thm_sublinear} disappears and (sublinear) communication does not improve the Stein-exponent, irrespective of the DMC's channel transition law. 
\end{corollary} 

\begin{IEEEproof}
The result only requires proof for DMCs satisfying  \eqref{eq:channel_cond0}, for which  Theorem~\ref{thm_sublinear} states that:
		\begin{IEEEeqnarray}{rCl}
\theta^\star_{\textnormal{sublin}}(\epsilon) &=&{\min_{\substack{\pi_{UV} \colon \\
\pi_U=P_U\\
\pi_V=P_V}}} D( \pi_{UV} \| Q_{UV}) \\
&=& {\min_{\substack{\pi_{UV} \colon \\
\pi_U=P_U\\
\pi_V=P_V}}} D( \pi_V \| Q_V) + \mathbb{E}_{\pi_V} [D( \pi_{U|V} \| Q_{U|V})] \IEEEeqnarraynumspace  \\
&\geq & {\min_{\substack{\pi_{UV} \colon \\
\pi_U=P_U\\
\pi_V=P_V}}} D( \pi_V \| Q_V) = D(P_V\| Q_V),
		\end{IEEEeqnarray}
		where the inequality holds 
with equality if, and only if, $\pi_{U|V}=Q_{U|V}$ is a permissible choice in the minimization. Notice that this is the case if, and only if, Condition \eqref{eq:cond} holds. 
\end{IEEEproof}

  \section{Distributed Detection with Sublinear Resources }
  
  \begin{figure}[t!]
\vspace{5mm}
\begin{center}
\includegraphics[scale=0.28]{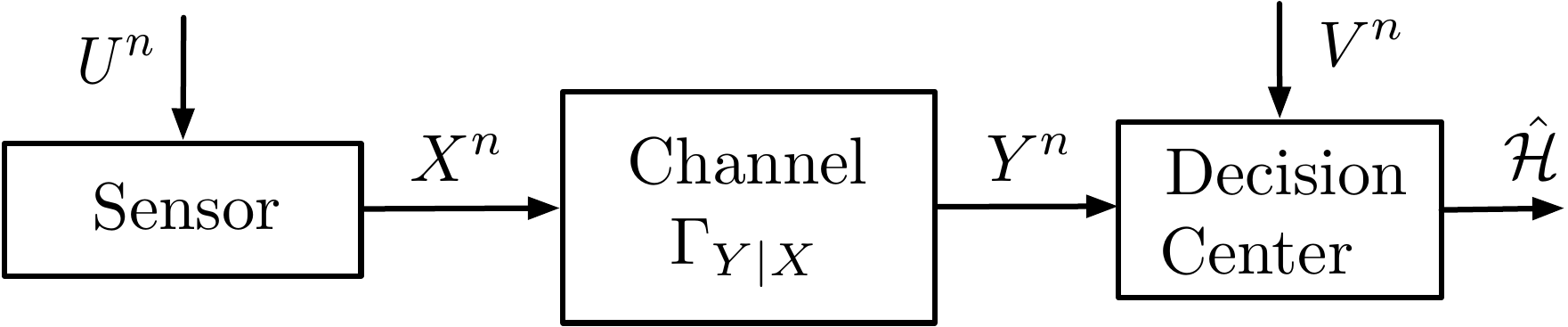}
\end{center}
\caption{System setup with a linear number of channel uses.}
\label{fig:DHT_linear}
\vspace{-3mm}
\end{figure}

We consider  almost the same setup, but  now communication takes place over $n$ channel uses (see Figure~\ref{fig:DHT_linear}) and a  stringent, sublinear block-power constraint is imposed. Encoder and decoder are  formalized by two functions $f^{(n)}$ and $g^{(n)}$ generating inputs and decisions as:
\begin{IEEEeqnarray}{rCl}X^n &=&f^{(n)}(U^n) \in \mathcal{X}^n\\
\label{eq:guess2}
\hat{\mathcal{H}}& =&	g^{(n)}(V^n,Y^n) \in\{0,1\}.
\end{IEEEeqnarray} 	

The encoding function is required to produce inputs satisfying a stringent resource constraint, described by a bounded and nonnegative cost function $c(\cdot)\colon \mathcal{X} \to \mathbb{R}_0^+$, for which we assume that there exists a unique zero-symbol:\begin{IEEEeqnarray}{rCl}
0 & \in& \mathcal{X} \\
 c(x)&=&0 \quad \textnormal{ if, and only if, }x=0.
 \end{IEEEeqnarray} The stringent resource constraint is described by  condition 
\begin{equation}\label{eq:cost}
\sum_{i=1}^n c(X_i) \leq C_n, \textnormal{ with prob. }1, 
\end{equation} 
for a given sequence
 $\{C_n\}$ that grows sublinearly in $n$:
\begin{subequations}\label{eq:cost_constraints}
\begin{IEEEeqnarray}{rCl}
	\lim_{n\to \infty} C_n &= &\infty \\
	\lim_{n\to \infty}\frac{C_n}{n}&= &0. 
\end{IEEEeqnarray} 	
\end{subequations}

 \begin{definition}\label{def:ach} Given  $\epsilon\in[0,1)$, a miss-detection error exponent $\theta>0$ is called $\epsilon$-achievable \emph{under stringent resource constraints $\{C_n\}$} if there exists a  sequence of encoding and decision functions   $\{(f^{(n)},g^{(n)})\}_{n=1}^\infty$  satisfying \eqref{eq:cost} and \eqref{eq:criteria2}. 
The supremum over all  miss-detection error exponents $\theta$ that are  $\epsilon$-achievable \emph{under stringent resource constraints $\{C_n\}$} is denoted $\theta_{\textnormal{str-cost}}^\star(\epsilon, \{C_n\})$.
 \end{definition}

As the following theorem shows, the largest type-II error exponent, i.e., the Stein-exponent, coincides exactly with the Stein-exponent in the setup where only a sublinear number of channel uses can be used for communication. Notice that the  resource constraint \eqref{eq:cost}  imposes that non-zero inputs can only be sent in a sublinear number of channel uses. However, the exact set  of channel uses that are used for transmission is now a design parameter, whereas in the previous section there was no freedom on the choice of channel uses that are effectively used for transmission. As our following result shows, this freedom in design cannot be used to increase the Stein-exponent. 

\medskip 

	\begin{theorem}\label{thm:resources}
	For any DMC $\Gamma_{Y|X}$, and sequence of cost-constraints $\{C_n\}$ satisfying \eqref{eq:cost_constraints}, we have
	\begin{equation}\label{th3}
	\theta_{\textnormal{str-cost}}^\star(\epsilon, \{C_n\}) = \theta_{\textnormal{sublin}}^\star(\epsilon), \quad \forall \epsilon \in[0,1).
	\end{equation}
\end{theorem}

\begin{IEEEproof}
Achievability follows directly from Theorem~\ref{thm_sublinear} and communicating only over the first $k'(n) = \frac{C_n}{ \min_{x \neq 0} c(x)}$ channel uses, which grows sublinearly in $n$. The converse is proved in  Appendix~\ref{app:2}.
\end{IEEEproof}

\section{Expected Resource Constraints}

Reconsider the same setup and definitions from the previous section, but where the almost-sure cost constraint \eqref{eq:cost} is replaced by an expected cost constraint:
\begin{equation}\label{eq:expected_cost}
\sum_{i=1}^n \mathbb{E}[c(X_i) |\mathcal{H}=\mathsf{H}]\leq C_n.
\end{equation} 
We shall impose above expected cost constraint either under both hypotheses or only under the null hypothesis $\mathcal{H}=0$. 

We denote the Stein-exponents by $\theta_{\textnormal{Exp-cost}, \mathcal{H}=0}^\star(\epsilon, \{C_n\})$ when constraint \eqref{eq:expected_cost} is applied only for $\mathsf{H}=0$ and by $\theta_{\textnormal{Exp-cost, both} }^\star(\epsilon, \{C_n\})$ when it is applied under both hypotheses.

To express our results in this section, we define the following three exponents: 
\begin{subequations}
\begin{IEEEeqnarray}{rCl}
E_1 & \triangleq &  \min_{\substack{\pi_{UV} \colon \\
\pi_U=P_U\\
\pi_V=P_V}} D( \pi_{UV} \| Q_{UV}) \\
E_2 & \triangleq &D( P_{V} \| Q_{V}) + \max_{\hat{x} \in \mathcal{X}\backslash \{0\}}  \hspace{-2mm} D( \Gamma_{Y|X=0}\| \Gamma_{Y|X=\hat{x}}) \label{eq:max4b}\\[1.2ex]
E_3 & \triangleq &  \min_{\substack{\pi_{UV} \colon \\
\pi_U=Q_U\\
\pi_V=P_V}} D( \pi_{UV} \| Q_{UV}) , 
\end{IEEEeqnarray} 
\end{subequations}
where notice that the difference between $E_1$ and $E_3$ only lies in the  minimization constraints $\pi_U=P_U$ or $\pi_U=Q_U$.

\begin{theorem} \label{thm:expected}
Fix $\epsilon \in [0,1)$ and a sequence of cost-constraints $\{C_n\}$ satisfying \eqref{eq:cost_constraints}. 
\begin{enumerate}
\item If the DMC is such that there exist two inputs $x_0, x_1\in\mathcal{X}$ and an output $y^*\in \mathcal{Y}$ satisfying the Conditions \eqref{eq:channel_cond0},
then
\begin{IEEEeqnarray}{rCl}\label{eq:IIIa}
\theta_{\textnormal{Exp-cost,both}}^\star(\epsilon, \{C_n\}) =\theta_{\textnormal{Exp-cost}, \mathcal{H}=0}^\star(\epsilon, \{C_n\}) &=& E_1. \IEEEeqnarraynumspace
\end{IEEEeqnarray}
\item Otherwise, if no triple $(x_0,x_1,y^*)$ satisfies \eqref{eq:channel_cond0}: 
\begin{IEEEeqnarray}{rCl}\label{eq:IIIb}
\theta_{\textnormal{Exp-cost}, \mathcal{H}=0}^\star(\epsilon,  \{C_n\}) &=& \min\big\{   E_1, E_2 \big \},
\end{IEEEeqnarray}
and  
\begin{IEEEeqnarray}{rCl}\label{eq:IIIc}
\theta_{\textnormal{Exp-cost,both}}^\star(\epsilon,  \{C_n\}) & = & \min\big\{   E_1,E_2, E_3 \big \}.\label{eq:IIId}
 \end{IEEEeqnarray}
\end{enumerate}	
\end{theorem}
\begin{IEEEproof} We obviously have the inequalities
\begin{IEEEeqnarray}{rCl}
\theta_{\textnormal{str-cost}}^\star(\epsilon,  \{C_n\}) \leq \theta_{\textnormal{Exp-cost,both}}^\star(\epsilon,  \{C_n\})  
 \end{IEEEeqnarray}
 and 
 \begin{IEEEeqnarray}{rCl}
\theta_{\textnormal{Exp-cost}, \mathcal{H}=0}^\star(\epsilon,  \{C_n\}) \geq \theta_{\textnormal{Exp-cost,both}}^\star(\epsilon,  \{C_n\}) . 
 \end{IEEEeqnarray}
 This implies directly the achievability proof for \eqref{eq:IIIa} and the converse in \eqref{eq:IIIc}. We thus have to prove the achievability results in \eqref{eq:IIIb} and \eqref{eq:IIId} and the converses in \eqref{eq:IIIa} and \eqref{eq:IIIb}. The achievability results are proved in Appendix~\ref{sec:achievability} and the converse results in Appendix \ref{sec:converse}. 
 \end{IEEEproof}

We observe that when Condition~\eqref{eq:channel_cond0} is satisfied, then the Stein-exponent coincides with the exponent reported for the previously studied steup. So, relaxing the  cost constraints to only hold on expectations does not increase the Stein-exponent. 

For fully-connected DMCs where any input can generate any output (and thus \eqref{eq:channel_cond0} is not satisfied), relaxing the stringent sublinear cost constraints to only hold on expectation however allows to achieve larger Stein-exponents. In fact, it allows to send high-cost inputs (with all non-zero symbols) with vanishing  probability. In our coding schemes in Appendix~\ref{sec:achievability}, we use this feature to send the zero inputs whenever the observation at the sensor is typical according to $P_U$ (and also when it is typical according to $Q_U$ in case the cost constraint is imposed also under $\mathcal{H}=1$) but it sends all non-zero inputs for all other observations. In this case, as our result shows, the obtained Stein-exponent also depends on the observations at the sensor and even on the best error-exponent that can be achieved when sending a single bit over the DMC. 

We conclude that under expected cost constraints, the observations at the sensor help increasing the Stein-exponent even when the cost constraint is sublinear in the blocklength and the DMC is fully-connected. Moreover, we can also observe that when $P_U\neq P_V$, then it depends on whether the expected resource constraint is imposed only under $\mathcal{H}=0$ or also under $\mathcal{H}=1$.

\section{Summary} 

This article presents a dichotomy in distributed detection under zero-rate and highly resource-limited communication over discrete memoryless channels (DMCs), based on the structure of the channel's transition law. When the DMC is fully-connected (i.e.,  every output symbol can be induced  by any input with positive probability), the maximum achievable Stein-exponent (largest type-II error exponent) collapses compared to cases where the channel is not fully-connected, i.e. only partially-connected.

In the partially-connected case, we showed that  the Stein-exponent matches that of a zero-rate noiseless link, in all three  communication settings that we consider: 1) the number of channel uses of the DMC is sublinear in the observation length $n$; 2) the number of channel uses is $n$ but a block-input cost constraint $C_n$ is imposed with probability 1 and $C_n$ is sublinear in $n$; 3) the block-input constraint $C_n$ is imposed only on expectation over the source observations. 

In contrast,  when the DMC is fully-connected, then we showed for cases 1) and 2) that the Stein-exponent degrades to that of a purely local test. In these cases, the remote sensor provides no benefit and the decision center should rely only on its own observations. Situation improves in case 3), where  communication from the sensor can improve performance and the decision center should base its decision on both its local observations and the outputs of the DMC. The optimal strategy for the sensor is to transmit zero symbols for typical source observations and non-zero symbols for atypical ones. We distinghuished two subcases of 3), where the expected cost constraint is either imposed under both hypotheses or just under $\mathcal{H}=0$. We determined the Stein-exponent for both subcases, and showed that they differ in general, except when the local observations at the sensor follow the same distributions under both marginals.

\section*{Acknowledgments}
The authors acknowledge funding from the ERC under Grant Agreement 101125691 and from the PEPR ``Networks of the Future" Initiative.

\appendices

\section{Converse Proof of Theorem~\ref{thm_sublinear}, Equation~\eqref{eq:Ib} }\label{app2}
 To prove the converse, we fix a sequence of encoding and decision functions $\{ f^{(n)},g^{(n)}\}_{n=1}^\infty$ such that $\varlimsup_{n\to \infty} \alpha_n \leq \epsilon$. To analyze the  type-II error probability of such a scheme, we introduce the notions of acceptance regions: 
\begin{equation}
\mathcal{A}_{V}(y^k) \triangleq \{ v^n \in \mathcal{V}^n \colon g^{(n)}( v^n, y^k)=0\}, \quad y^k \in \mathcal{Y}^k,
\end{equation}
and 
\begin{equation}
\mathcal{A}_{V} \triangleq \bigcup_{y^k\in \mathcal{Y}^k} \mathcal{A}_{V}(y^k).
\end{equation}
We can then write the miss-detection error probability as:
\begin{IEEEeqnarray}{rCl}
\lefteqn{\beta_n} \nonumber\\
&=&\Pr\left[\hat{\mathcal{H}}=0 |\mathcal{H}=1\right]\\
& = & \sum_{y^k} \Pr\left[ Y^k=y^k, V^n \in \mathcal{A}_V(y^k)|\mathcal{H}=1\right] \\
& = & \sum_{y^k} \sum_{v^n \in \mathcal{A}_V(y^k)}   \sum_{u^n} \Pr\left[ Y^k=y^k, V^n=v^n, U^n=u^n|\mathcal{H}=1\right]  \nonumber \\\\
& = & \sum_{y^k} \sum_{v^n \in \mathcal{A}_V(y^k)}   \sum_{u^n}  \Pr\left[V^n=v^n, U^n=u^n|\mathcal{H}=1\right]   \nonumber \\
&& \hspace{2cm} \cdot\Pr\left[ Y^k=y^k| U^n=u^n \right].
\end{IEEEeqnarray}
We continue to bound the second probability, by noticing that 
\begin{equation}
\Pr\left[ Y^k=y^k| U^n=u^n \right]  \geq \gamma_{\min}^k,
\end{equation}
where we define 
\begin{equation}
\gamma_{\min} \triangleq \min_{x,y} \Gamma_{Y|X}(y|x),
\end{equation}
which is strictly positive by assumption. 
Thus: 
\begin{IEEEeqnarray}{rCl}
\lefteqn{\beta_n} \nonumber \\
& \geq &  \gamma_{\min}^k  \sum_{y^k} \sum_{v^n \in \mathcal{A}_V(y^k)}   \sum_{u^n}  \Pr\left[V^n=v^n, U^n=u^n|\mathcal{H}=1\right] \IEEEeqnarraynumspace \\
&=&   \gamma_{\min}^k  \sum_{y^k} \sum_{v^n \in \mathcal{A}_V(y^k)}    \Pr\left[V^n=v^n|\mathcal{H}=1\right] \\
&\geq &\gamma_{\min}^k  \cdot  \Pr[ V^n \in \mathcal{A}_V |\mathcal{H}=1].
\end{IEEEeqnarray}
Since $k$ grows sublinearly in $n$, we obtain that the type-II error probability of the chosen  encoding and decision functions is bounded by 
\begin{IEEEeqnarray}{rCl}
\varlimsup_{n\to \infty} - \frac{1}{n} \log \beta_n \leq \varlimsup_{n\to \infty} - \frac{1}{n} \log  \Pr[ V^n \in \mathcal{A}_V |\mathcal{H}=1], \IEEEeqnarraynumspace
\end{IEEEeqnarray}
and thus is bounded by the type-II error exponent of a local test at the decision center with acceptance region $\mathcal{A}_V$.

Notice next that under $\mathcal{H}=0$, the $V^n$ sequence falls in $\mathcal{A}_V$ with probability at least 
\begin{equation}
  \Pr[ V^n \in \mathcal{A}_V |\mathcal{H}=0] \geq    \Pr[ V^n \in  \mathcal{A}_V(Y^k) |\mathcal{H}=0] = 1- \alpha_n, 
\end{equation} 
and thus, by assumption,  the type-I error probability of the local test on $V^n$ with acceptance region $\mathcal{A}_V$ satisfies
  \begin{IEEEeqnarray}{rCl}
 \varlimsup_{n\to \infty}  \Pr[ V^n \notin \mathcal{A}_V |\mathcal{H}=0] &\leq &   \epsilon <1. 
  \end{IEEEeqnarray}
We can now invoke the standard Stein lemma, which states that the type-II error probability of any local test on $V^n$ with type-I error probability bounded away from 1 satisfies 
  \begin{IEEEeqnarray}{rCl}
  \varlimsup_{n\to \infty} - \frac{1}{n} \log  \Pr[ V^n \in \mathcal{A}_V |\mathcal{H}=1] \leq D(P_V\|Q_V),
  \end{IEEEeqnarray}
  which concludes the proof. 
  
  \section{Achievability Proof of Theorem \ref{thm_sublinear}, Equation  \eqref{eq:Ia}}\label{app:1}
We recall the proposed coding scheme in Section \ref{theorem1}. 


%
%

Define  $\gamma_{x_0} \triangleq\Gamma_{Y|X}(y^*|x_0)$.

\noindent\underline{Analysis of $\alpha_n$:}
\begin{IEEEeqnarray}{rCl}
\lefteqn{1-\alpha_n} \nonumber\\
&=&\Pr\left[\hat{\mathcal{H}}=0 |\mathcal{H}=0\right]\\
&=&\Pr\left[ \exists i\in\{1,\ldots, k\} \colon Y_i=y^* \; \textnormal{and} \; V^n \in \mathcal{T}^{(n)}_{\mu}(P_{V}) |\mathcal{H}=0\right] \nonumber\\  \\
&\stackrel{(a)}{=} & \Pr\big[ \exists i\in\{1,\ldots, k\} \colon Y_i=y^* \; \textnormal{and} \; V^n \in \mathcal{T}^{(n)}_{\mu}(P_{V})\;  \textnormal{and} \nonumber\\
&&\hspace{2cm}  X^k=x_0^k \; \textnormal{and} \; U^n \in \mathcal{T}^{(n)}_{\mu}(P_{U}) \big|\mathcal{H}=0\big] 		\IEEEeqnarraynumspace\\
&=& \Pr\big[ V^n \in \mathcal{T}^{(n)}_{\mu}(P_{V})\; \textnormal{and} \;  U^n \in \mathcal{T}^{(n)}_{\mu}(P_{U}) |\mathcal{H}=0\big] \nonumber\\
&& \cdot \underbrace{\Pr\big[ X^k=x_0^k | U^n \in \mathcal{T}^{(n)}_{\mu}(P_{U})\big] }_{=1}\nonumber\\
&& \cdot \Pr\big[ \exists i\in\{1,\ldots, k\} \colon Y_i=y^* |X^k=x_0^k,  \mathcal{H}=0\big]\\
&= & \Pr\big[ V^n \in \mathcal{T}^{(n)}_{\mu}(P_{V})\; \textnormal{and} \;  U^n \in \mathcal{T}^{(n)}_{\mu}(P_{U}) |\mathcal{H}=0\big] \nonumber\\
&&  \cdot  (1- (1-\gamma_{x_0})^k),\label{eq:a}
\end{IEEEeqnarray}
where $(a)$ holds because the output symbol $y^*$ can occur from input $x_0$ but not from input $x_1$ and because input $X^k=x_0^k$ is sent only if  $U^n \in \mathcal{T}^{(n)}_{\mu}(P_{U})$. 

Since $\gamma_{x_0}$ lies in the half-open interval $(0,1]$, we have $(1-\gamma_{x_0})^{k(n)}$ that tends to 0 as $n\to \infty$. Moreover, by the weak law of large numbers, irrespective of $\mu>0$:
\begin{IEEEeqnarray}{rCl}
\lim_{n\to \infty}  \Pr\big[ V^n \in \mathcal{T}^{(n)}_{\mu}(P_{V})\; \textnormal{and} \;  U^n \in \mathcal{T}^{(n)}_{\mu}(P_{U})\big |\mathcal{H}=0\big] =1. \nonumber\\
\end{IEEEeqnarray}
Plugging these limits into \eqref{eq:a}, we can conclude that the type-I error probability of our scheme vanishes: 
\begin{IEEEeqnarray}{rCl}
\lim_{n\to \infty} \alpha_n=0.
\end{IEEEeqnarray}

\noindent\underline{Analysis of $\beta_n$ and $\theta$:}
Similarly to above, we have:
\begin{IEEEeqnarray}{rCl}
\beta_n
&=&\Pr\left[\hat{\mathcal{H}}=0 |\mathcal{H}=1\right]\\
&{=} & \Pr\big[ \exists i\in\{1,\ldots, k\} \colon Y_i=y^* \; \textnormal{and} \; V^n \in \mathcal{T}^{(n)}_{\mu}(P_{V})\nonumber\\
&&\hspace{1.2cm}  \textnormal{and} \; X^k=x_0^k \; \textnormal{and} \; U^n \in \mathcal{T}^{(n)}_{\mu}(P_{U}) |\mathcal{H}=1\big] 		\IEEEeqnarraynumspace\\
&=& \Pr\big[ V^n \in \mathcal{T}^{(n)}_{\mu}(P_{V})\; \textnormal{and} \;  U^n \in \mathcal{T}^{(n)}_{\mu}(P_{U}) |\mathcal{H}=1\big] \nonumber\\
&& \cdot \underbrace{\Pr\big[ X^k=x_0^k | U^n \in \mathcal{T}^{(n)}_{\mu}(P_{U})\big] }_{=1}\nonumber\\
&& \cdot \Pr\big[ \exists i\in\{1,\ldots, k\} \colon Y_i=y^* |X^k=x_0^k|\mathcal{H}=1\big]\\
&\leq & \Pr\big[ V^n \in \mathcal{T}^{(n)}_{\mu}(P_{V})\; \textnormal{and} \;  U^n \in \mathcal{T}^{(n)}_{\mu}(P_{U}) |\mathcal{H}=1\big]  \\
&\leq & (n+1)^{|\mathcal{U}||\mathcal{V}|} 2^{-n \min D( \pi_{UV}\|Q_{UV})} \label{eq:b}
\end{IEEEeqnarray}

We can conclude that 
\begin{IEEEeqnarray}{rCl}
\varliminf_{n\to \infty} -\frac{1}{n} \log \beta_n \geq \min D( \pi_{UV}\|Q_{UV}),
\end{IEEEeqnarray}
where the minimum is now over all pmfs $ \pi_{UV}\in\mathcal{P}(\mathcal{U}\times \mathcal{V})$ with marginals satisfying $| \pi_U(u) - P_U(u)| \leq \mu$ and $| \pi_V(u) - P_V(u)| \leq \mu$. Picking $\mu>0$ sufficiently small, all type-II error exponents smaller than the right-hand side of \eqref{eq:Ia} can be shown to be achievable. 
\medskip

\section{Converse Proof of Theorem~\ref{thm:resources}}\label{app:2}


To prove the two converses to \eqref{eq:Ia} and \eqref{eq:Ib}, reuse the definition 
\begin{IEEEeqnarray}{rCl}\label{eq:def_kprime}
k'(n)&\triangleq & \frac{C_n}{c_{\min}},
\end{IEEEeqnarray} 	
where $c_{\min}$ denotes the minimum cost of any non-zero input: 
\begin{equation}
c_{\min} \triangleq \min_{ x\neq 0} c(x)
\end{equation} 
with $c_{\min}>0$. Notice that  $k'(n)$ bounds the number of non-zero symbols in each possible  input sequence $x^n$. The channel input sequence $X^n$ thus lies with probability~1 in the set of all low-weight inputs
\begin{equation}
\tilde{\mathcal{X}}^n := \{ x^n \in \mathcal{X}^n \colon w_{\textnormal{H}}(x^n) \leq  k'(n)\}.
\end{equation}
  We show in the following that the cardinality of $\tilde{\mathcal{X}}^n$ grows sublinearly in $n$, which immediately establishes the converse result to \eqref{eq:Ia} because the type-II error exponent in this setup cannot exceed the type-II error exponent in the setup of \cite{Shalaby} where communication consists in sending a sublinear number of noiseless bits.

To see that  the cardinality of $\tilde{\mathcal{X}}^n$ grows sublinearly in $n$, notice that this set  can  be described as the union over all  type-classes (i.e., sets of sequences  with same type) for types that assign frequency larger or equal to  $1-\frac{k'(n)}{n}$ to the 0 symbol. 
Since  the type-class for type $\boldsymbol{\pi}$ is of size at most $2^{n H(\boldsymbol{\pi})}$ and  the number of type-classes is bounded by $(n+1)^{|\mathcal{X}|}$, we have: 
\begin{IEEEeqnarray}{rCl}
 \big|\tilde{\mathcal{X}}^n \big| &\leq&  (n+1)^{|\mathcal{X}|} 2^{n \max_{\boldsymbol{\pi}} H( \boldsymbol{\pi})},
\end{IEEEeqnarray} 
where the maximum is over all types $\boldsymbol{\pi}$ with $\boldsymbol{\pi}(0)\geq  1-\frac{k'(n)}{n}$.  Since $k'(n)$ grows sublinearly in $n$
and by the continuity of the entropy functional, we  obtain 
\begin{IEEEeqnarray}{rCl}\label{eq:cardinality_limit}
\lefteqn{\lim_{n\to \infty} \frac{1}{n} \log \big| \tilde{\mathcal{X}}^n \big|} \nonumber\\
&\leq & \lim_{n\to \infty} \left[ \frac{|\mathcal{X}|}{n} \log (n+1) + \max_{ \substack{\boldsymbol{\pi} \colon \\ \boldsymbol{\pi}(0) \geq 1-\frac{k'(n)}{n}}}  H( \boldsymbol{\pi}) \right] =0. \IEEEeqnarraynumspace
\end{IEEEeqnarray}

To prove the converse to \eqref{eq:Ib}, we notice that for any two input sequences $x_1^n, x_2^n \in \tilde{\mathcal{X}}^n$ and any output sequence $y^n \in \mathcal{Y}^n$ we have:
\begin{equation}
\gamma_{Q}^{2 k'(n)} \leq \frac{\Pr\left[ Y^n=y^n|X^n=x_1^n\right]}{\Pr\left[ Y^n=y^n| X^n=x_2^n \right] }  \leq \gamma_{Q}^{-2 k'(n)},
\end{equation}
where 
\begin{equation}\label{eq:quot}
\gamma_{Q} \triangleq \min_{ \substack{x_1,x_2,y \colon\\ x_1\neq x_2}} \frac{\Gamma_{Y|X}(y|x_1)}{\Gamma_{Y|X}(y|x_2)}  \in(0,1]. 
\end{equation}
Trivially, this implies also the bounds:
\begin{equation}\label{eq:bound_ch}
\gamma_{Q}^{2 k'(n)} \leq \frac{\Pr\left[ Y^n=y^n|U^n=u_1^n\right]}{\Pr\left[ Y^n=y^n| U^n=u_2^n \right] }  \leq \gamma_{Q}^{-2 k'(n)},
\end{equation}
for any two input sequences $u_1^n, u_2^n \in {\mathcal{U}}^n$ and  output sequence $y^n \in \mathcal{Y}^n$, and in particular for any two measures $R$ and $R'$ on $\mathcal{U}^n$:
\begin{IEEEeqnarray}{rCl}\label{eq:in}
\lefteqn{\mathbb{E}_R  \left[ \sum_{y^n}\Pr\left[ Y^n=y^n|U^n\right]  \right] } \nonumber \\
& \geq & \gamma_{Q}^{2 k'(n)} \mathbb{E}_{R'}\left[ \sum_{y^n} \Pr\left[ Y^n=y^n| U^n \right] \right].
\end{IEEEeqnarray} 


Define  the acceptance regions
\begin{equation}
\mathcal{A}_{V}(y^n) \triangleq \{ v^n \in \mathcal{V}^n \colon g^{(n)}( v^n, y^n)=0\}, \quad y^n \in \mathcal{Y}^n.
\end{equation}
Similarly to the converse proof to \eqref{eq:Ib}---but where $y^k$ needs to be replaced by $y^n$---we have: 
\begin{IEEEeqnarray}{rCl}
\beta_n
& = & \sum_{y^n} \sum_{v^n \in \mathcal{A}_V(y^n)}   \Pr\left[V^n=v^n|\mathcal{H}=1\right]   \nonumber \\
&& \hspace{1cm} \cdot\Pr\left[ Y^n=y^n| V^n=v^n , \mathcal{H}=1\right] \vspace{2mm}\\
& \stackrel{(a)}{\geq } & \sum_{y^n} \sum_{v^n \in \mathcal{A}_V(y^n)}  \Pr\left[V^n=v^n |\mathcal{H}=1\right]   \nonumber \\
&& \hspace{1cm} \cdot\Pr\left[ Y^n=y^n|  V^n=v^n , \mathcal{H}=0\right]  \gamma_{Q}^{2 k'(n)}
 \label{eq:prelim}
\end{IEEEeqnarray}
where {Inequality} $(a)$ holds by Inequality \eqref{eq:in} because both  $\Pr\left[ Y^n=y^n| V^n=v^n , \mathcal{H}=1\right]$ and $\Pr\left[ Y^n=y^n| V^n=v^n , \mathcal{H}=0\right]$ are expectations  of the probabilities $\Pr\left[ Y^n=y^n| U^n \right]$ with respect to the two measures $P_{U|V}^{\otimes n}$ and $Q_{U|V}^{\otimes n}$.

By assumption,
\begin{equation}
\lim_{n\to \infty} -\frac{1}{n} \log   \gamma_{Q}^{2 k'(n)} = 0, 
\end{equation}
and thus 
\begin{IEEEeqnarray}{rCl}\label{eq:A}
\lefteqn{\varlimsup_{n\to \infty} -\frac{1}{n} \log \beta_n } \nonumber \;\; \\
& \leq & \varlimsup_{n\to \infty} -\frac{1}{n} \log \sum_{y^n} \sum_{v^n \in \mathcal{A}_V(y^n)}   \ \Pr\left[V^n=v^n |\mathcal{H}=1\right]   \nonumber \\
&& \hspace{3cm} \cdot \Pr\left[ Y^n=y^n|  V^n=v^n , \mathcal{H}=0\right]. \IEEEeqnarraynumspace
\end{IEEEeqnarray}
\begin{figure}[t!]
\begin{center}
\includegraphics[scale=0.27]{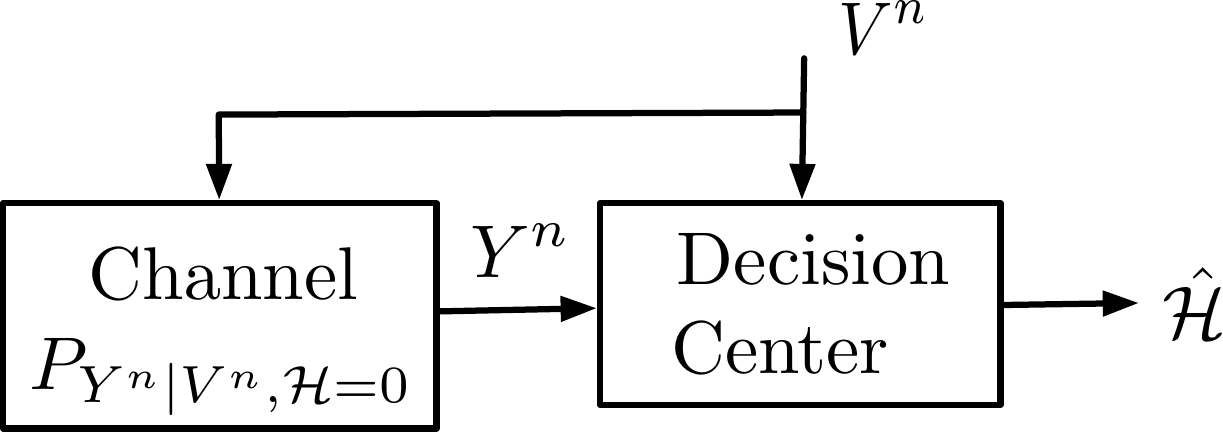}
\end{center}
\caption{Derived binary hypothesis test with channel $P_{Y^n|V^n, \mathcal{H}=0}$ used under both hypotheses.}
\label{fig:HT_derived}
\vspace{-3mm}

\end{figure}
Notice however that the right-hand side of the above equation corresponds to the miss-detection error exponent of a local detection problem of the form in Figure~\ref{fig:HT_derived}, where the decision center observes $V^n$, which is i.i.d. according to $P_V$ or $Q_V$ depending on the two hypotheses,  and  the outcome ${Y}^n$, which is generated from $V^n$ according to the law $ \Pr\left[ Y^n=y^n|  V^n=v^n , \mathcal{H}=0\right]$ irrespective of the  hypothesis. This hypothesis testing problem is however a special case of a randomized decision rule for a binary hypothesis test with observation $V^n$. Moreover, we know that under $\mathcal{H}=0$ for the randomized acceptance region $\mathcal{A}_{V}(Y^n)$:
\begin{IEEEeqnarray}{rCl}
1-\alpha_n& \triangleq & \sum_{y^n} \sum_{v^n \in \mathcal{A}_V(y^n)}   \ \Pr\left[V^n=v^n |\mathcal{H}=0\right]   \nonumber \\
&& \hspace{1cm} \cdot \Pr\left[ Y^n=y^n|  V^n=v^n , \mathcal{H}=0\right], 
\end{IEEEeqnarray}
which by assumption is bounded from below by $1-\epsilon>0$. 
By standard Stein-exponent arguments, we  can thus deduce  that the Limit on the right-hand side of \eqref{eq:A} is bounded above by $D(P_V \| Q_V)$,
which concludes the proof.

 \section{Achievability Proof of Theorem \ref{thm:expected}}\label{sec:achievability}
 \subsection{Proof of Achievability for Equation \eqref{eq:IIIb}} 
Consider the following scheme. Let $\hat{x}$ be the maximizer in \eqref{eq:max4b} and define
\begin{equation}
\mu_n  \triangleq \sqrt{\frac{|\mathcal{U}|c(\hat{x})}{ 4C_n}} . 
\end{equation}
Notice that by Assumptions \eqref{eq:cost_constraints}: 
\begin{subequations}\label{eq:mu_eq}
\begin{IEEEeqnarray}{rCl}
\lim_{ n\to \infty} \mu_n &=&0 \\
\lim_{ n\to \infty} \mu_n^2 n &=&\infty.
 \end{IEEEeqnarray}
 \end{subequations}
\medskip 
\noindent\underline{Sensor:} If $U^n \in \mathcal{T}_{\mu_n}(P_{U})$,  send $X^n=0^n$. Else, send $X^n=\hat{x}^n$. 

\noindent\underline{Decision Center:} If $Y^{n} \in \mathcal{T}_{\mu_n}^{(n)}(\Gamma_{Y|X}(\cdot|x=0))$ and  $V^n \in \mathcal{T}_{\mu}^{(n)}(P_{V})$,  declare $\hat{\mathcal{H}}=0$. Otherwise, declare $\hat{\mathcal{H}}=1$. 

\noindent\underline{Analysis of cost constraint:} The expected cost is: 
\begin{IEEEeqnarray}{rCl}
\sum_{i=1}^n \mathbb{E}[c(X_i) |\mathcal{H}=0]&= & \Pr[ U^n \notin  \mathcal{T}_{\mu_n}^{(n)}(P_{U}) |\mathcal{H}=0] \cdot n  c(\hat{x})\nonumber\\
 \\
& \leq & \frac{ |\mathcal{U}|}{4 \mu_n^2 n} n c(\hat{x}) = C_n, 
 \end{IEEEeqnarray}
where in the inequality we used the well-known inequality on the probability of the typical set from \cite[Remark to Lemma 2.12]{csiszar_book}. 

\noindent\underline{Analysis of $\alpha_n$:} We have: 
\begin{IEEEeqnarray}{rCl}
\lefteqn{1-\alpha_n}\nonumber\\
&=&\Pr\left[\hat{\mathcal{H}}=0 |\mathcal{H}=0\right]\label{eq:01}\\
&=&\Pr\left[ Y^n \in  \mathcal{T}_{\mu_n}^{(n)}(\Gamma_{Y|X}(\cdot|0)) \; \textnormal{and} \; V^n \in \mathcal{T}_{\mu}^{(n)}(P_{V}) |\mathcal{H}=0\right] \nonumber\\
 \\
&\geq & \Pr\big[ Y^n \in  \mathcal{T}_{\mu_n}^{(n)}(\Gamma_{Y|X}(\cdot|0)) \; \textnormal{and} \; \nonumber \\
&& \hspace{1.4cm} U^n \in  \mathcal{T}_{\mu_n}^{(n)}(P_{U})\; \textnormal{and} \; V^n \in \mathcal{T}_{\mu}^{(n)}(P_{V}) |\mathcal{H}=0\big]\IEEEeqnarraynumspace\\ 
& =  & \Pr[ Y^n \in  \mathcal{T}_{\mu_n}^{(n)}(\Gamma_{Y|X}(\cdot|0)) |X^n=0^n]  \nonumber \\
&&  \cdot\Pr\left[ U^n \in  \mathcal{T}_{\mu_n}^{(n)}(P_{U})\; \textnormal{and} \; V^n \in \mathcal{T}_{\mu_n}^{(n)}(P_{V}) |\mathcal{H}=0\right] .\label{eq:fin}
\end{IEEEeqnarray}
Notice that by the weak law of large numbers and because the sequence $\mu_n$ satisfies \eqref{eq:mu_eq}, both probabilities in \eqref{eq:fin} tend to 1 and thus
\begin{IEEEeqnarray}{rCl}
\lim_{n\to \infty} \alpha_n=0.
\end{IEEEeqnarray}

\noindent\underline{Analysis of $\beta_n$ and $\theta$:}
Similarly to above, we have:
\begin{IEEEeqnarray}{rCl}
\beta_n&=&\Pr\left[\hat{\mathcal{H}}=0 |\mathcal{H}=1\right]\\
& = &  \Pr\Big[ Y^n \in  \mathcal{T}_{\mu_n}^{(n)}(\Gamma_{Y|X}(\cdot|0)) \; \textnormal{and} \; \nonumber \\
&& \hspace{1.4cm} U^n \in  \mathcal{T}_{\mu_n}^{(n)}(P_{U})\; \textnormal{and} \; V^n \in \mathcal{T}_{\mu_n}^{(n)}(P_{V}) \; \big|\;\mathcal{H}=1\Big] \nonumber \\
&& +  \Pr\Big[ Y^n \in  \mathcal{T}_{\mu_n}^{(n)}(\Gamma_{Y|X}(\cdot|0)) \; \textnormal{and} \;\nonumber \\
&& \hspace{1.4cm} U^n \notin  \mathcal{T}_{\mu_n}^{(n)}(P_{U})\; \textnormal{and} \; V^n \in \mathcal{T}_{\mu_n}^{(n)}(P_{V}) \; \big|\;\mathcal{H}=1\Big] \nonumber \\\\
& \leq  & \Pr\left[ U^n \in  \mathcal{T}_{\mu_n}^{(n)}(P_{U})\; \textnormal{and} \; V^n \in \mathcal{T}_{\mu_n}^{(n)}(P_{V}) \; \big|\;\mathcal{H}=1\right] \nonumber\\
&& +  \Pr\left[ Y^n \in  \mathcal{T}_{\mu_n}^{(n)}(\Gamma_{Y|X}(\cdot|0))\; \big|\;X^n=\hat{x}^n\right]  \nonumber \\
&&  \cdot\Pr\left[  V^n \in \mathcal{T}_{\mu_n}^{(n)}(P_{V}) |\mathcal{H}=1\right] \\
&\leq &     (n+1)^{|\mathcal{U}||\mathcal{V}|} 2^{-n \min D( \pi_{UV}\|Q_{UV})} \nonumber\\ 
&&+  (n+1)^{|\mathcal{Y}|} 2^{-n \min D( \pi_{Y}\|\Gamma_{Y|X=\hat{x}})}  \nonumber \\
&& \cdot    (n+1)^{|\mathcal{V}|} 2^{-n \min D( \pi_{V}\|Q_{V})} ,
\end{IEEEeqnarray}
where the three  minima are over  types  $\pi_{UV}$, $\pi_Y$, and $\pi_V$ satisfying the respective typicality constraints. We next notice that since $\mu_n\to 0$ as $n\to \infty$, the following limits hold as $n\to \infty$: 
\begin{IEEEeqnarray}{rCl}
\lim_{n\to \infty}  \min D( \pi_{Y}\|\Gamma_{Y|X=\hat{x}})& =& D(\Gamma_{Y|X=0} \| \Gamma_{Y|X=\hat{x}}) \IEEEeqnarraynumspace\\
\lim_{n\to \infty}  \min D( \pi_{V}\|Q_{V})&= & D(P_V\| Q_V) \\
\lim_{n\to \infty}  \min D( \pi_{UV}\|Q_{UV})&= & E_1.  
\end{IEEEeqnarray}
We can thus conclude that for the proposed scheme: 
\begin{IEEEeqnarray}{rCl}
\lim_{n\to \infty} -\frac{1}{n} \log \beta_n \geq   \min\{E_1, E_2\}. 
\end{IEEEeqnarray}
\smallskip

 \subsection{Proof of Achievability for Equation \eqref{eq:IIId}} 
The scheme is similar to the scheme in the previous section, except that the sensor also sends $X^n=0^n$ when $U^n \in \mathcal{T}_{\mu_n}^{(n)}(Q_{U})$. 

\noindent\underline{Analysis of cost constraint:} The expected cost  under hypothesis $\mathcal{H}=0$ satisfies 
\begin{IEEEeqnarray}{rCl}
\sum_{i=1}^n \mathbb{E}[c(X_i) |\mathcal{H}=0]& \leq  & \Pr[ U^n \notin  \mathcal{T}_{\mu_n}^{(n)}(P_{U})  |\mathcal{H}=0] \cdot n  c(\hat{x})\IEEEeqnarraynumspace \\
& \leq & \frac{ |\mathcal{U}|}{4 \mu_n^2 n} n c(\hat{x}) = C_n, 
 \end{IEEEeqnarray}
 and similarly, under $\mathcal{H}=1$: 
\begin{IEEEeqnarray}{rCl}
\sum_{i=1}^n \mathbb{E}[c(X_i) |\mathcal{H}=1]& \leq  & \Pr[ U^n \notin  \mathcal{T}_{\mu_n}^{(n)}(Q_{U})|\mathcal{H}=1 ] \cdot n  c(\hat{x})\IEEEeqnarraynumspace \\
& \leq & \frac{ |\mathcal{U}|}{4 \mu_n^2 n} n c(\hat{x}) = C_n, 
 \end{IEEEeqnarray}
 
\noindent\underline{Analysis of $\alpha_n$:} One can follow exactly the same analysis steps \eqref{eq:01}--\eqref{eq:fin}  as in the previous section. 

\noindent\underline{Analysis of $\beta_n$:}
We have:
\begin{IEEEeqnarray}{rCl}
\lefteqn{\beta_n} \nonumber \\
&=&\Pr\left[\hat{\mathcal{H}}=0 |\mathcal{H}=1\right]\\
& = &  \Pr\big[ Y^n \in  \mathcal{T}_{\mu_n}^{(n)}(\Gamma_{Y|X}(\cdot|0)) \; \textnormal{and} \;  V^n \in \mathcal{T}^{(n)}_{\mu}(P_{V})\nonumber \\
&& \hspace{3cm}  \textnormal{and} \; U^n \in  \mathcal{T}_{\mu_n}^{(n)}(P_{U}) |\mathcal{H}=1\big] \nonumber \\
&  &+   \Pr\big[ Y^n \in  \mathcal{T}_{\mu_n}^{(n)}(\Gamma_{Y|X}(\cdot|0))\; \textnormal{and} \; V^n \in \mathcal{T}^{(n)}_{\mu}(P_{V})  \nonumber \\
&& \hspace{3cm} \; \textnormal{and} \;U^n \in  \mathcal{T}_{\mu_n}^{(n)}(Q_{U}) |\mathcal{H}=1\big] \nonumber \\
&& +  \Pr\Big[ Y^n \in  \mathcal{T}_{\mu_n}^{(n)}(\Gamma_{Y|X}(\cdot|0)) \; \textnormal{and} \; V^n \in \mathcal{T}_{\mu}^{(n)}(P_{V})\nonumber \\
&& \hspace{2cm}\; \textnormal{and} \; U^n \notin ( \mathcal{T}_{\mu_n}^{(n)}(P_{U}) \cup \mathcal{T}_{\mu_n}^{(n)}(Q_{U}) )  |\mathcal{H}=1\Big] \nonumber\\
& \leq  & \Pr\left[ U^n \in  \mathcal{T}_{\mu_n}^{(n)}(P_{U})\; \textnormal{and} \; V^n \in \mathcal{T}_{\mu}^{(n)}(P_{V}) |\mathcal{H}=1\right] \nonumber\\
&&+  \Pr\left[ U^n \in  \mathcal{T}_{\mu_n}^{(n)}(Q_{U})\; \textnormal{and} \; V^n \in \mathcal{T}_{\mu}^{(n)}(P_{V}) |\mathcal{H}=1\right] \nonumber\\
&& +  \Pr[ Y^n \in  \mathcal{T}_{\mu_n}^{(n)}(\Gamma_{Y|X}(\cdot|0)) |X^n=\hat{x}^n]  \nonumber \\
&& \qquad  \cdot\Pr\left[  V^n \in \mathcal{T}_{\mu}^{(n)}(P_{V}) |\mathcal{H}=1\right] \\
&\leq &   (n+1)^{|\mathcal{U}||\mathcal{V}|} 2^{-n \min D( \pi_{UV}\|Q_{UV})}  \nonumber\\
& & + (n+1)^{|\mathcal{U}||\mathcal{V}|} 2^{-n \min D( \pi_{UV}\|Q_{UV})}  \nonumber\\
& &+2^{-n \min D( \pi_{Y}\|\Gamma_{Y|X=\hat{x}})}     (n+1)^{|\mathcal{V}|+|\mathcal{Y}|} 2^{-n \min D( \pi_{V}\|Q_{V})} ,\nonumber\\
\end{IEEEeqnarray}
where  the first two minimima are  over  types  $\pi_{UV}$ with marginals  that are $\mu_n$-close to $P_U$ and $P_V$ or to $Q_U$ and $P_V$, respectively,  and the latter two minima are over types $\pi_Y$ and $\pi_V$ that are $\mu_n$-close to $\Gamma_{Y|X=0}$ and $P_V$, respectively.  Letting $n\to \infty$, by standard arguments we have: 
\begin{IEEEeqnarray}{rCl}
\varliminf_{n\to \infty} -\frac{1}{n} \log \beta_n \geq  \min\{E_1, E_2, E_3\}. 
\end{IEEEeqnarray}
\smallskip

 \section{Converse Proof for Theorem \ref{thm:expected}} \label{sec:converse}
The following considerations will be used in all our converse proofs. 
Define $k'(n)$ an increasing function of positive integers  satisfying: 
\begin{subequations}\label{eq:kprime}
\begin{IEEEeqnarray}{rCl}
	\varliminf_{n\to \infty} \frac{k'(n)}{C_n}&= &\infty \label{eq:lim_I_kprime} \\
	\lim_{n\to \infty}\frac{k'(n)}{n}&= &0.  \label{eq:lim_II_kprime}
\end{IEEEeqnarray} 	
\end{subequations}
Also, define the set of $u^n$-sequences that are mapped to ``low-weight inputs"
\begin{equation}
\tilde{\mathcal{U}}^n := \{ u^n \in \mathcal{U}^n \colon w_{\textnormal{H}}(f^{(n)}(u^n)) \leq k'(n)\}, 
\end{equation}
Under the stringent resource constraint, we have: 
\begin{IEEEeqnarray}{rCl}
C_n & \geq &\sum_{i=1}^n \mathbb{E}[  c(X_i) | \mathcal{H}=\mathsf{H}]    \\
&\geq  & c_{\min}\cdot  \mathbb{E}[ w_{\textnormal{H}}(X^n) | \mathcal{H}=\mathsf{H}] \\
 & \geq  & c_{\min}\cdot k'(n) \cdot  \Pr[U^n \notin \tilde{\mathcal{U}}^n | \mathcal{H}=\mathsf{H}], 
\end{IEEEeqnarray} 
where  $c_{\min} =\min_{ x\neq 0} c(x) >0$.
Then, the following simple bound holds: 
\begin{IEEEeqnarray}{rCl}\label{eq:k}
\Pr[U^n\notin \tilde{\mathcal{U} }| \mathcal{H}=\mathsf{H}] \leq \frac{C_n}{c_{\min}\cdot k'(n)}.
\end{IEEEeqnarray}
By our assumptions on $k'(n)$ in \eqref{eq:lim_I_kprime}, the set $\tilde{\mathcal{U}}$ thus occurs asymptotically with probability~$1$:
\begin{IEEEeqnarray}{rCl}\label{eq:prob1}
\lim_{n\to \infty} \Pr\left[U^n \notin \tilde{\mathcal{U}}^n\; \Big | \mathcal{H}=\mathsf{H}\right]  &= &0, 
\end{IEEEeqnarray} 
under any of the hypotheses $\mathsf{H}$ for which the 
stringent resource constraint is  imposed. 

Let $\{\mu_n\}_n$ be any sequence satisfying
\begin{subequations}\label{eq:mun}
\begin{IEEEeqnarray}{rCl}
\lim_{ n\to \infty} \mu_n &=&0 \\
\lim_{ n\to \infty} \mu_n^2 n &=&\infty.
 \end{IEEEeqnarray}
 \end{subequations}
Since the probability of the set $\mathcal{T}^{(n)}_{\mu_n/2}(P_U)$ asymptotically tends to 1 as $n\to \infty$ under hypothesis $\mathsf{H}=0$,  we conclude that if Limit~\eqref{eq:prob1} holds for  $\mathsf{H}=0$, then 
\begin{equation} 
\lim_{n \to \infty}\Pr\left[ U^n \in \left(\mathcal{T}^{(n)}_{\mu_n/2}(P_U) \cap \tilde{\mathcal{U}}^n \right)\; \Big | \; \mathcal{H}=0 \right] =1.\label{eq:lim1}
\end{equation} 

Similarly, we can conclude that if Limit~\eqref{eq:prob1} holds for $\mathsf{H}=1$, then 
\begin{equation} \label{eq:dd5}
\lim_{n \to \infty}\Pr\left[ U^n \in \left(\mathcal{T}^{(n)}_{\mu_n/2}(Q_U) \cap \tilde{\mathcal{U}}^n \right)\;\Big | \; \mathcal{H}=1 \right] =1,
\end{equation} 
and because all sequences in $\mathcal{T}^{(n)}_{\mu_n/2}(Q_U)$ have almost equal probability: 
\begin{IEEEeqnarray}{rCl}
\lim_{n \to \infty}\frac{ \left| \mathcal{T}^{(n)}_{\mu_n/2}(Q_U) \cap  \tilde{\mathcal{U}}^n  \right| }{ \left| \mathcal{T}^{(n)}_{\mu_n/2}(Q_U)\right|  } &=& 1.\label{eq:allQU}
\end{IEEEeqnarray}

 \subsection{Upper Bounds $E_1$ and $E_3$}
 
Upper bounds $E_1$ and $E_3$ can be proved using similar change of measure arguments.

\noindent\underline{\textit{The region $\mathcal{F}_{n}$ and its probability:}}
Let  $F_{UV}$ be an arbitrary pmf with marginals $F_V=P_V$ and $F_U=P_U$ if the expected cost constraint is only imposed under $\mathcal{H}=0$, and marginals $F_V=P_V$ and $F_U \in \{P_U, Q_U\}$ if the expected cost constraint  is imposed under both hypotheses.

 Define the region 
\begin{IEEEeqnarray}{rCl}
	{\mathcal{F}}_{n}&\triangleq & \{ (u^n,v^n,y^n) \colon (u^n,v^n) \in \mathcal{T}_{\mu_n}^{(n)}(F_{UV}) ,  \nonumber\\
	&&\hspace{1.3cm}\; u^n \in (\tilde{\mathcal{U}}^n \cap  \mathcal{T}_{\mu_n/2}^{(n)}(F_{U}))  ,   \; g(v^n,y^n)=0 \}.\nonumber\\
\end{IEEEeqnarray}
For the probability of this set under $\mathcal{H}=0$ we have: 
\begin{IEEEeqnarray}{rCl}
\Delta_n&\triangleq &\Pr\left[  (U^n,V^n, Y^n) \in \mathcal{F}_n | \mathcal{H}=0 \right]  \\
& = &   \Pr\Big [ U^n \in ( \tilde{\mathcal{U}}^n \cap \mathcal{T}^{(n)}_{\mu_n/2}(F_U))  | \mathcal{H}=0 \Big] \nonumber\\
&  &\cdot \Pr\Big[  (U^n,V^n) \in \mathcal{T}_{\mu_n}^{(n)}(F_{UV}) |   U^n \in ( \tilde{\mathcal{U}}^n \cap \mathcal{T}^{(n)}_{\mu_n/2}(F_U)),  \nonumber\\ 
&& \hspace{5.3cm} \mathcal{H}=0 \Big] \nonumber\\
& & \cdot \Pr\Big[   g^{(n)}(V^n,Y^n)=0| U^n \in (\tilde{\mathcal{U}}^n \cap \mathcal{T}_{\mu_n/2}^{(n)}(F_U)),\nonumber\\
&& \hspace{2.7cm}  (U^n,V^n) \in \mathcal{T}_{\mu_n}^{(n)} (F_{UV}) ,  \mathcal{H}=0 \Big]. \nonumber\\\label{eq:three} 
\end{IEEEeqnarray}
We bound each of these probabilities individually. Notice first that by \eqref{eq:prob1} and \eqref{eq:lim1} for $F_U=P_U$ we have:
\begin{IEEEeqnarray}{rCl}
 \Pr\Big [ U^n \in ( \tilde{\mathcal{U}}^n \cap \mathcal{T}^{(n)}_{\mu_n/2}(P_U))  | \mathcal{H}=0 \Big] 
&\geq  &1- o(1). \IEEEeqnarraynumspace\label{eq:f1}
\end{IEEEeqnarray}
If \eqref{eq:allQU} holds, then because all sequences of $ \mathcal{T}^{(n)}_{\mu_n/2}(F_U))$ have approximately equal probability, we have for  $F_U=Q_U$: 
\begin{IEEEeqnarray}{rCl}
\Pr\Big [ U^n \in ( \tilde{\mathcal{U}}^n \cap \mathcal{T}^{(n)}_{\mu_n/2}(F_U))  | \mathcal{H}=0 \Big]
&\geq  & 2^{-n ( D(F_U \| P_U) +o(1))}.  \nonumber\\\label{eq:f2}
\end{IEEEeqnarray}

We further observe that by well-known arguments, if $F_{V|U}\neq P_{V|U}$:
\begin{IEEEeqnarray}{rCl}
\lefteqn{\Pr\Big[  (U^n,V^n) \in \mathcal{T}_{\mu_n}^{(n)}(F_{UV}) |   U^n \in ( \tilde{\mathcal{U}}^n \cap \mathcal{T}^{(n)}_{\mu_n/2}(F_U)), \; \mathcal{H}=0 \Big]} \nonumber \\
&\geq & 2^{-n (D(  F_{UV} \| F_UP_{V|U})+o(1))}. \hspace{4.8cm}\label{eq:f3} 
\end{IEEEeqnarray}
If $F_{V|U}=P_{V|U}$:
\begin{IEEEeqnarray}{rCl}
\lefteqn{\Pr\Big[  (U^n,V^n) \in \mathcal{T}_{\mu_n}^{(n)}(F_{UV}) |   U^n \in ( \tilde{\mathcal{U}}^n \cap \mathcal{T}^{(n)}_{\mu_n/2}(F_U)), \; \mathcal{H}=0 \Big]} \nonumber \\
&\geq & 1- o(1). \hspace{7.3cm}\label{eq:f4} 
\end{IEEEeqnarray}

It remains to analyze the third probabilty in \eqref{eq:three}. For ease of notation, let $T_{U^nV^n}(u^n,v^n)$ denote the pmf of the tuples $(u^n, v^n)$ conditional on $\mathcal{H}=0$ and on the conditions $(U^n,V^n) \in \mathcal{T}_{\mu_n}^{(n)}(F_{UV})$ and $U^n \in (\tilde{\mathcal{U}}^n \cap \mathcal{T}_{\mu_n/2}^{(n)}(F_U))$: 
\begin{IEEEeqnarray}{rCl}
\lefteqn{T_{U^nV^n}(u^n,v^n) } \; \nonumber\\
&:=& \Pr[U^n=u^n, V^n=v^n| U^n \in (\tilde{\mathcal{U}}^n \cap \mathcal{T}_{\mu_n/2}^{(n)}(F_U)),  \nonumber \\
&& \hspace{2.6cm} (U^n,V^n) \in \mathcal{T}_{\mu_n}^{(n)}(F_{UV}),  \mathcal{H}=0].\IEEEeqnarraynumspace
\end{IEEEeqnarray} 
Similarly, let 
 $S_{U^nV^n}(u^n,v^n)$ denote the pmf of the tuples $(u^n, v^n)$ conditional on $\mathcal{H}=0$ and on the condition $U^n \in \tilde{\mathcal{U}}^n$: 
\begin{IEEEeqnarray}{rCl}
S_{U^nV^n}(u^n,v^n)
&:=& \Pr[U^n=u^n, V^n=v^n| U^n \in \tilde{\mathcal{U}}^n,  \mathcal{H}=0]. \nonumber\\
\end{IEEEeqnarray}
By the definition of the set $\tilde{\mathcal{U}}^n$, we can write:
\begin{IEEEeqnarray}{rCl}
\lefteqn{ \Pr\Big[   g^{(n)}(V^n,Y^n)=0| U^n \in (\tilde{\mathcal{U}}^n \cap \mathcal{T}_{\mu_n/2}^{(n)}(F_U)), }\nonumber\\
&& \hspace{3cm}  (U^n,V^n) \in \mathcal{T}_{\mu_n}^{(n)}(F_{UV}) ,  \mathcal{H}=0 \Big]\nonumber \IEEEeqnarraynumspace\\
&= & \sum_{\substack{y^n}}  \sum_{\substack{v^n \in A_V(y^n)}} \nonumber\\
&& \hspace{.5cm}  \cdot   \sum_{\substack{u^n \in \tilde{\mathcal{U}}^n}} \Pr[Y^n=y^n| U^n=u^n]  \cdot T_{U^nV^n}(u^n,v^n) \nonumber\\\\
&\geq & \sum_{\substack{y^n}}  \sum_{\substack{v^n \in A_V(y^n)}} \nonumber\\
&& \cdot   \gamma_{Q}^{2 k'(n)}  \sum_{\substack{u^n \in \tilde{\mathcal{U}}^n}} \Pr[Y^n=y^n| U^n=u^n]  S_{U^nV^n}(u^n,v^n)\nonumber\\ \\
& = & \gamma_{Q}^{2 k'(n)}  \Pr\Big[   g^{(n)}(V^n,Y^n)=0| U^n \in \tilde{\mathcal{U}}^n,  \mathcal{H}=0 \Big],
\end{IEEEeqnarray}
where the inequality holds because $u^n \in \tilde{\mathcal{U}}^n$ and thus the corresponding inputs all have weights bounded by $k'(n)$, see also \eqref{eq:in} and  the definition of $\gamma_Q$ in \eqref{eq:quot}. 

Since for any events $A$ and $B$ it holds that $\Pr[A|B]\geq \Pr[A \cap B] = \Pr[A]- \Pr[A \cap B^{c}] \geq  \Pr[A]- \Pr[B^{c}]$, we can further bound: 
\begin{IEEEeqnarray}{rCl}
\lefteqn{ \Pr\Big[   g^{(n)}(V^n,Y^n)=0| U^n \in \tilde{\mathcal{U}}^n ,  \mathcal{H}=0 \Big]  }\; \nonumber \\
& \geq & \Pr\Big[   g^{(n)}(V^n,Y^n)=0 | \mathcal{H}=0 \Big]  - \Pr\Big[ U^n \notin \tilde{\mathcal{U}}^n |  \mathcal{H}=0 \Big] \nonumber\\ \\
& \geq  & 1- \epsilon - o(1) ,
\end{IEEEeqnarray} 
where the last inequality holds by the assumption on the type-I error probability  and by \eqref{eq:lim1}.

Combining all these observations, we obtain:
\begin{itemize} 
\item \textit{If $F_U=P_U$ and $F_{V|U}=P_{V|U}$}, 
\begin{IEEEeqnarray}{rCl}
\varlimsup_{n\to \infty} -\frac{1}{n} \log \Delta_n =0.
\end{IEEEeqnarray}

\item \textit{If $F_U=P_U$ and $F_{V|U}\neq P_{V|U}$}, 
\begin{IEEEeqnarray}{rCl}
\varlimsup_{n\to \infty} -\frac{1}{n} \log \Delta_n \leq D(F_{UV}\| P_{U}P_{V|U}).
\end{IEEEeqnarray}

\item \textit{If $F_U=Q_U$ and $F_{V|U}=P_{V|U}$}, 
\begin{IEEEeqnarray}{rCl}
\varlimsup_{n\to \infty} -\frac{1}{n} \log \Delta_n \leq D(Q_{U}\| P_{U})=D(F_{UV}\| P_{U}P_{V|U}).\nonumber\\
\end{IEEEeqnarray}
\item \textit{If $F_U=Q_U$ and $F_{V|U}\neq P_{V|U}$}
\begin{IEEEeqnarray}{rCl}
\varlimsup_{n\to \infty} -\frac{1}{n} \log \Delta_n& \leq& D(Q_{U}\| P_{U}) +D(F_{UV}\| Q_{U}P_{V|U})\nonumber\\ \\
&=&D(F_{UV}\| P_{U}P_{V|U}).
\end{IEEEeqnarray}
\end{itemize} 
So, in all cases we have: 
\begin{IEEEeqnarray}{rCl}\label{eq:Delta_nlimit}
\varlimsup_{n\to \infty} -\frac{1}{n} \log \Delta_n \leq D(F_{UV}\| P_{UV}).
\end{IEEEeqnarray}\smallskip

\noindent\underline{\textit{Change of measure on $\mathcal{F}_n$:}}\\
Define now the tuple $(\tilde{U}^n,\tilde{V}^n,\tilde{Y}^n)$ to be of joint pmf
\begin{IEEEeqnarray}{rCl}\label{eq:norm}
P_{\tilde{U}^n\tilde{V}^n\tilde{Y}^n} (u^n,{v}^n,y^n) & = & \frac{P_{{UV}}^{\otimes n}(u^n,{v}^n) \Gamma_{Y|X}^{\otimes n} ( y^n | f^{(n)}(u^n) )}{\Delta_n}  \nonumber \\
&& \cdot \mathds{1}\{ (u^n,v^n,y^n) \in \mathcal{F}_n\}. \IEEEeqnarraynumspace
\end{IEEEeqnarray}

We obtain:
\begin{IEEEeqnarray}{rCl}
\lefteqn{ -\frac{1}{n} \log \beta_n }  \nonumber \\
 & = &  -\frac{1}{n} \log \Pr[ \hat{\mathcal{H}}=0 | \mathcal{H}=1]  \nonumber \\
&= &  -\frac{1}{n} \log \left(\! \sum_{\substack{u^n, v^n, y^n:\\ g(v^n, y^n)=0}}\!\!\!\!\! Q_{UV}^{\otimes n}(u^n, v^n) \cdot \Pr[ Y^n=y^n | U^n = u^n] \right) \nonumber \\ 
& & \\
&\leq &  -\frac{1}{n} \log \left( \sum_{\substack{(u^n, v^n, y^n) \\ \in \mathcal{F}_n}} \!\!\!\!\! Q_{UV}^{\otimes n}(u^n, v^n) \cdot \Pr[ Y^n=y^n | U^n = u^n] \right) \nonumber \\
\\ 
 & = &  -\frac{1}{n} \log Q_{U^nV^nY^n}\left(  {\mathcal{F}_n} \right)   \\
 & \stackrel{(a)}{ =}  &  \frac{1}{n} D\left( P_{\tilde{U}^n\tilde{V}^n\tilde{Y}^n}\left(  {\mathcal{F}_n} \right) \| Q_{U^nV^nY^n}\left(  {\mathcal{F}_n} \right)\right) \label{eq:18}\\
 &\stackrel{(b)}{ \leq}  & \frac{1}{n} D\left( P_{\tilde{U}^n\tilde{V}^n\tilde{Y}^n} \| Q_{UV}^{\otimes n} P_{Y^n|U^n} \right)\\
&\stackrel{(c)}{ =} & \frac{1}{n} \sum_{(u^n,v^n)} P_{\tilde{U}^n\tilde{V}^n}(u^n, v^n) \log \frac{  P_{UV}^{\otimes n}(u^n, v^n)  }{ Q_{UV}^{\otimes n}(u^n,v^n)} \nonumber \\
&&- \frac{1}{n} \log \Delta_n   \\
& = &  \frac{1}{n}  \sum_{i=1}^n \sum_{(u^n,v^n)}P_{\tilde{U}^n\tilde{V}^n}(u^n, v^n)\log \frac{  P_{UV}(u_i, v_i)  }{ Q_{UV}(u_i,v_i)} \nonumber \\
&&- \frac{1}{n} \log \Delta_n  \\
& = &  \frac{1}{n}  \sum_{i=1}^n \sum_{ u_i,v_i}P_{\tilde{U}_i\tilde{V}_i}(u_i, v_i)\log \frac{  P_{UV}(u_i, v_i)  }{ Q_{UV}(u_i,v_i)} \nonumber \\
&&- \frac{1}{n} \log \Delta_n  \\
& = &  \sum_{u,v}P_{\tilde{U}_T\tilde{V}_T}(u,v) \log \frac{  P_{UV}(u, v)  }{ Q_{UV}(u,v)}  - \frac{1}{n} \log \Delta_n ,  \label{eq:lastbound}
\end{IEEEeqnarray}
where $T$ is a uniform random variable over $\{1,\ldots, n\}$ independent of $(\tilde{U}^n, \tilde{V}^n)$. 
In above inequalities, $(a)$ holds because $ P_{\tilde{U}^n\tilde{V}^n\tilde{Y}^n}$ is only defined on $\mathcal{F}_n$ and thus $ P_{\tilde{U}^n\tilde{V}^n\tilde{Y}^n}(\mathcal{F}_n)=1$; $(b)$ holds by the data processing inequality; 
and $(c)$ holds by the definition of $ P_{\tilde{U}^n\tilde{V}^n\tilde{Y}^n}$. 

We  use Limit~\eqref{eq:Delta_nlimit}  and the fact that $P_{\tilde{U}_T\tilde{V}_T}(u,v)=F_{UV}(u,v)+o(1)$ because $\mathcal{F}_n$ only contains $(u^n,v^n)$-sequences that are $F_{UV}$-typical. Then, taking  the limit $n\to \infty$ on \eqref{eq:lastbound},
we obtain:
\begin{IEEEeqnarray}{rCl}
\lefteqn{ \varlimsup_{n\to \infty} -\frac{1}{n} \log \beta_n }  \nonumber \\
&\leq &  \sum_{u,v}F_{UV} (u ,v) \log \frac{  P_{UV}(u, v)  }{ Q_{UV}(u,v)}  + D(F_{UV} \| P_{UV})  \IEEEeqnarraynumspace
\\ 
&=& D( F_{UV}\| Q_{UV}).\label{eq:r}
\end{IEEEeqnarray}

\noindent\underline{\textit{Concluding the Proof:}} 

Recall that we have derived the bound \eqref{eq:r} for any pmf $F_{UV}$ with marginal $F_V=P_V$ and   $F_U =P_U$ if the expected cost constraint is imposed only under hypothesis $\mathcal{H}=0$ and marginals  $F_V=P_V$ and $F_U\in \{P_U, Q_U\}$ if the cost constraint is imposed under both hypotheses.

By Limits \eqref{eq:lim1} and~\eqref{eq:r}, we thus obtain  
 the desired converse result $E_1$ on both $\theta_{\textnormal{Exp-cost}, \mathcal{H}=0}$ and  $\theta_{\textnormal{Exp-cost}, \textnormal{both}}$, and by  also using Limit~\eqref{eq:dd5}, we establish the converse result $E_3$ on $\theta_{\textnormal{Exp-cost}, \textnormal{both}}$. 
\medskip

\subsection{Upper Bound $E_2$} 
We again use change of measure arguments to obtain upper bound $E_2$, but now combined with the blowing up lemma. 


Define  for each sequence $u^n$, the set of indices where this sequence induces input zero: 
\begin{equation}
\mathcal{I}_{u^n}\triangleq \{i \in\{1,\ldots, n\} \colon f_i(u^n)=0\}
\end{equation}
for $f_i(u^n)$ the $i$-th component of $f^{(n)}(u^n)$.

Let $\mathcal{A}_n$ denote the acceptance region:
\begin{equation}
\mathcal{A}_n \triangleq \{ (v^n,y^n) \colon g(v^n, y^n)=0\}.
\end{equation} 
Further, let $\{\ell_n\}$ be a sequence with $\lim_{n\to\infty}\ell_n/\sqrt{n}=\infty$ and $\lim_{n\to\infty}\ell_n/n=0$. Define then for each $v^n$  the $\ell_n$-blown-up: acceptance region
\begin{IEEEeqnarray}{rCl}
	\hat{\mathcal{A}}_n^{\ell_n}(v^n)\triangleq \left\{ \tilde{y}^n\colon \exists y^n\in\mathcal{A}_n(v^n,y^n)\;\;\text{s.t.}\;\;d_{\text{H}}(\tilde{y}^n,y^n)\leq \ell_n \right\}.\nonumber\\
\end{IEEEeqnarray}

We next define the set on which we will be employing the change of measure. Fix $\eta \in (0, (1-\epsilon))$ and define
\begin{IEEEeqnarray}{rCl}
\mathcal{D}_n &\triangleq  & \{ (u^n, v^n) \colon   u^n \in \tilde{\mathcal{U}} ,\   v^n \in \mathcal{T}_{\mu_n}^{(n)}(P_V), \nonumber \\
 && \hspace{1.3cm}  \Pr[ (v^n,Y^n) \in \mathcal{A}_n | U^n=u^n, V^n=v^n] \geq \eta \}, \nonumber\\
\end{IEEEeqnarray}
where notice that  the last probability is not conditioned on the hypothesis as  it only depends on the encoding function and the channel transition law. 

We continue to apply the union bound to obtain:
\begin{IEEEeqnarray}{rCl}
\Delta_n& \triangleq &\Pr \left[ (U^n,V^n) \in \mathcal{D}_n \, \big | \, \mathcal{H}=0 \right]\\
&\geq & 1- \Pr[ U \notin \tilde{\mathcal{U}}^n| \, \mathcal{H}=0] -  \Pr[  v^n \notin \mathcal{T}_{\mu_n}^{(n)}(P_V) |\mathcal{H}=0] \nonumber\\
&& - \Pr[ \;  \Pr[ (V^n,Y^n) \in \mathcal{A}_n | U^n, V^n]   < \eta ]\IEEEeqnarraynumspace\\
& \geq& 1 - \frac{C_n}{c_{\min}k'(n)} - \frac{ |\mathcal{V}|}{4 \mu_n^2 n} - \frac{1-\epsilon}{\eta} \label{eq:D_nbound}.\IEEEeqnarraynumspace
\end{IEEEeqnarray}

To prove the desired converse result, we define the new pmf
\begin{IEEEeqnarray}{rCl}
\lefteqn{\tilde{P}_{U^nV^nY^n\mathcal{I}_{U^n}}(u^n,v^n,y^n,\mathcal{I}) }\nonumber\\
 &=  &\frac{P_{UV}^{\otimes n}(u^n, v^n)}{\Delta_n} \cdot \mathbbm{1}\{ (u^n, v^n) \in \mathcal{D}_n\} \cdot \Gamma_{Y|X}^{\otimes n} (y^n | {f}^{(n)}(u^n)) \nonumber \\
&& \cdot \mathbbm{1}\{ \mathcal{I} =\mathcal{I}_{u^n}\}
\end{IEEEeqnarray}
and notice that this new measure satisfies: 
\begin{IEEEeqnarray}{rCl}
\lefteqn{\tilde{P}(Y^n\in \hat{\mathcal{A}}^{\ell_n}(V^n))} \nonumber\\
&  = & \sum_{(u^n,v^n)\in \mathcal{D}_n}  \tilde{P}_{U^nV^n}(u^n,v^n) \nonumber\\
& & \hspace{1cm}\cdot  \Pr[ Y^n \in \hat{\mathcal{A}}^{\ell_n}(v^n) | U^n=u^n,V^n=v^n]  \nonumber\\
& \geq &   \sum_{(u^n,v^n)\in \mathcal{D}_n}  \tilde{P}_{U^nV^n}(u^n,v^n) \left(1-\frac{\sqrt{n\ln 1/\eta}}{\ell_n}\right) \nonumber\\
& =& 1- \lambda_n,\label{eq:P0}
\end{IEEEeqnarray}
where we defined $\lambda_n \triangleq \frac{\sqrt{n\ln 1/\eta}}{\ell_n}$, which by our assumptions tends to 0 for $n \to \infty$; and the inequality is obtained  by applying
  the blowing-up lemma \cite[remark p. 446]{MartonBU} on each pair $(u^n,v^n)$. 
  
  We next examine how the blow-up of  the acceptance region influences the type-II error probability: 
  \begin{IEEEeqnarray}{rCl}
\lefteqn{Q\left(Y^n \in \hat{\mathcal{A}}^{\ell_n}(V^n)\right)} \qquad \nonumber\\
&  \triangleq& \sum_{(u^n,v^n)}  Q_{UV}^{\otimes n}(u^n,v^n) \nonumber\\
& & \hspace{.2cm}\cdot  \Pr[ Y^n \in \hat{\mathcal{A}}^{\ell_n}(v^n) | U^n=u^n,V^n=v^n]  \\[1.2ex]
&\leq & \sum_{(u^n,v^n)}  Q_{UV}^{\otimes n}(u^n,v^n) \nonumber\\
& & \hspace{1cm}\cdot  \Pr[ Y^n \in \mathcal{A}(v^n) | U^n=u^n,V^n=v^n]  \nonumber\\
&& \hspace{1cm} \cdot e^{n h_{\textnormal{b}}(\ell_n/n)} \cdot|\mathcal{Y}|^{\ell_n}\cdot \gamma_{\min}^{\ell_n} \label{step62}\\
	&=&\beta_n\cdot  e^{n h_{\textnormal{b}}(\ell_n/n)} \cdot|\mathcal{Y}|^{\ell_n}\cdot \gamma_{\min}^{\ell_n} ,\label{step64}
\end{IEEEeqnarray}
where recall that  $\gamma_{\min}\triangleq  \min_{(x,y)}\Gamma_{y|x}(y|x)$, and where the inequality can be obtaiend following the steps in   \cite[the Proof of Lemma 5.1]{Csiszarbook}.

Let now 
  \begin{IEEEeqnarray}{rCl}
\tilde{P}_{\mathcal{H}}(0) &\triangleq &\tilde{P}\left(Y^n\in\hat{\mathcal{A}}^{\ell_n}(V^n)\right) \\{
Q}_{\mathcal{H}}(0)&\triangleq &Q\left(Y^n\in\hat{\mathcal{A}}^{\ell_n}(V^n)\right)
\end{IEEEeqnarray} and  $\tilde{P}_{\mathcal{H}}(1)$ and $Q_{\mathcal{H}}(1)$ their complements. By the data processing inequality:
  \begin{IEEEeqnarray}{rCl}D(\tilde{P}_{V^nY^n} \|  Q_{V^nY^n}) &\geq & D(\tilde{P}_{\mathcal{H}} \|  Q_{\mathcal{H}}) \\
& \geq & -1 + \tilde{P}_{\mathcal{H}}(0) \log \frac{1}{ Q_{\mathcal{H}}(0)},\label{eq:D}
\end{IEEEeqnarray}
where the second inequality is obtained by writing out the divergence term and bounding binary entropy by 1. 

Combining  \eqref{eq:P0}  and \eqref{step64} with \eqref{eq:D}, we can then write: 
\begin{IEEEeqnarray}{rCl}
\lefteqn{- \frac{1}{n} \log \beta_n} \nonumber\\ 
& \leq& \frac{1}{(1- \lambda_n)n}\left(  1+ D\left( \tilde{P}_{V^nY^n} \| Q_{V^n Y^n}\right)\right) \nonumber\\
& & +h_{\textnormal{b}}(\ell_n/n) +\frac{\ell_n}{n}\log( \gamma_{\min}\cdot |\mathcal{Y}|) \\
&=&  \frac{1}{n(1-\lambda_n)}  \left( 1+ D\left( \tilde{P}_{V^n} \| Q^{\otimes n}_{V} \right) \right) \nonumber \\
&&+  \frac{1}{n(1 -\lambda_n)}\mathbb{E}_{\tilde{P}_{V^n}}\left[      D\left( \tilde{P}_{Y^n|V^n} \| Q_{Y^n|V^n} \right)\right] \Bigg)\nonumber\\
& & +h_{\textnormal{b}}(\ell_n/n) +\frac{\ell_n}{n}\log( \gamma_{\min}\cdot |\mathcal{Y}|),\label{eq:summands}
\end{IEEEeqnarray}
where the inequality holds by the data-processing inequality. 

Following previous arguments,  we have
\begin{IEEEeqnarray}{rCl}
 \lefteqn{\frac{1}{n}  D\left( \tilde{P}_{V^n} \| Q^{\otimes n}_{V} \right)  } \nonumber\\&\leq  & \frac{1}{n} \sum_{v^n}  \tilde{P}_{V^n}(v^n) \log \left( \frac{{P}_{V}^{\otimes n}(v^n) }{Q_{V}^{\otimes n}(v^n)}\right) -  \log \Delta_n , \label{eq:DV}
\end{IEEEeqnarray}
which tends to $D(P_V\| Q_V)$ by our definition of $\tilde{P}_{V^n}$ and because $\Delta$ is bounded away from 0 and smaller than 1, see \eqref{eq:D_nbound}.

Before  considering the second term in the decomposition \eqref{eq:summands}, we observe that by basic probability:
\begin{IEEEeqnarray}{rCl} 
\tilde{P}_{Y^n|V^n}(y^n|v^n) &=& \mathbb{E}_{\tilde{P}_{\mathcal{I}_{U^n} } } \left[ \tilde{P}_{Y^n|V^n\mathcal{I}_{U^n}}(y^n|v^n, \mathcal{I}_{U^n})  \right]\IEEEeqnarraynumspace \\
Q_{Y^n|V^n}(y^n|v^n) &=& \mathbb{E}_{\tilde{P}_{\mathcal{I}_{U^n} } }\left[ {Q}_{Y^n|V^n}(y^n|v^n)  \right] , 
\end{IEEEeqnarray}
and thus by the convexity of KL divergence: 
\begin{IEEEeqnarray}{rCl} 
\lefteqn{
  \frac{1}{n}\mathbb{E}_{\tilde{P}_{V^n}}\left[      D\left( \tilde{P}_{Y^n|V^n} \| Q_{Y^n|V^n} \right)\right] } \nonumber\\
  &\leq &   \frac{1}{n} \mathbb{E}_{\tilde{P}_{V^n\mathcal{I}_{U^n} } }\left[  D\left( \tilde{P}_{Y^n|V^n\mathcal{I}_{U^n}} \| Q_{Y^n|V^n} \right)\right].
  \end{IEEEeqnarray}
  Moreover, for given realizations $\mathcal{I}_{U^n}  = \mathcal{I}$ and $V^n=v^n$, we can decompose the divergence as:
\begin{IEEEeqnarray}{rCl} 
\lefteqn{  
    \frac{1}{n}D\left( \tilde{P}_{Y^n|V^n=v^n,\mathcal{I}_{U^n} = \mathcal{I}} \| Q_{Y^n|V^n=v^n} \right) } \nonumber\\
  & =&    \frac{1}{n}D\left( \tilde{P}_{Y_{\mathcal{I}}^n|V^n=v^n,\mathcal{I}_{U^n} =\mathcal{I} } \| Q_{Y^n_{\mathcal{I}}|v^n} \right) \nonumber\\
  & & + \frac{1}{n}\mathbb{E} \left[ D\left( \tilde{P}_{Y_{{\mathcal{I}}^c}^n|V^n=v^n,\mathcal{I}_{U^n} =\mathcal{I}} \| Q_{Y^n_{\mathcal{I}^c}|V^n=v^n} \right)  \right] \IEEEeqnarraynumspace \\
 &\leq &  \frac{1}{n} D\left( \tilde{P}_{Y_{\mathcal{I}}^n|V^n=v^n,\mathcal{I}_{U^n} =\mathcal{I}}\| Q_{Y^n_{\mathcal{I}}|V^n=v^n} \right) \nonumber \\  
 & & - \frac{n-| \mathcal{I}| }{n} \log \gamma_{\min}\IEEEeqnarraynumspace \\
  &=&  \frac{1}{n} D\left( \Gamma_{Y|X}(\cdot|0)^{\otimes |\mathcal{I}|}\| Q_{Y^n_{\mathcal{I}}|V^n=v^n} \right)   - \frac{n-  |\mathcal{I}| }{n} \log \gamma_{\min}.\label{eq:ne}\nonumber \\ 
    \end{IEEEeqnarray}

Notice that the second term in \eqref{eq:ne} vanishes as $n\to \infty$ by our assumption that $\gamma_{\min}$ is bounded away from zero and because under $\tilde{P}$ the number of non-zero input symbols is upper bounded by $C_{n}/c_{\min}$ and thus sublinear in $n$. 

    The first term in \eqref{eq:ne} can be bounded using the concavity of the $\log$-function:     
    \begin{IEEEeqnarray}{rCl} 
\lefteqn{   \frac{1}{n} D\left( \Gamma_{Y|X}(\cdot|0)^{\otimes |\mathcal{I}|}\| Q_{Y^n_{\mathcal{I}}|v^n} \right) } \nonumber\\
&\leq &  \frac{1}{n} \sum_{x^n} Q_{X^n|v^n}(x^n) D\left( \Gamma_{Y|X}(\cdot|0)^{\otimes |\mathcal{I}|}\| Q_{Y^n_{\mathcal{I}}|x^n_{\mathcal{I}}} \right) \\
&=& \frac{|\mathcal{I}|}{n} \sum_{x^n} Q_{X^n|v^n}(x^n)  \nonumber\\
&& \qquad  \sum_{x' \in \mathcal{X}} \mathbf{\pi}_{x^n}(x') D\left( \Gamma_{Y|X}(\cdot|0)\| \Gamma_{Y|X}(\cdot|x') \right)\IEEEeqnarraynumspace
\\
& \leq&   \sum_{x^n} Q_{X^n|v^n}(x^n) \max_{x' \in\mathcal{X}} D\left( \Gamma_{Y|X}(\cdot|0)\| \Gamma_{Y|X}(\cdot|x') \right)\IEEEeqnarraynumspace
\\
&= & \max_{x' \in\mathcal{X}} D\left( \Gamma_{Y|X}(\cdot|0)\| \Gamma_{Y|X}(\cdot|x') \right).
        \end{IEEEeqnarray}

Combining these observations with  \eqref{eq:summands} and \eqref{eq:DV} and using that $\ell_n/n \to 0$ as $n\to \infty$, we can conclude; 
    \begin{IEEEeqnarray}{rCl} \varlimsup_{n\to \infty} -\frac{1}{n} \log \beta_n &\leq &D(P_V\|Q_V) \nonumber\\
& & +  \max_{x' \in\mathcal{X}} D\left( \Gamma_{Y|X}(\cdot|0)\| \Gamma_{Y|X}(\cdot|x') \right),\IEEEeqnarraynumspace
        \end{IEEEeqnarray}which concludes the proof.

\bibliographystyle{ieeetr}
\bibliography{references}

\begin{thebibliography}{10}

\bibitem{Ahlswede}
R.~Ahlswede and I.~Csisz\'ar, ``Hypothesis testing with communication
  constraints,'' {\em IEEE Trans.~Inf.~Theory}, vol.~32, pp.~533--542, Jul.
  1986.

\bibitem{Han}
T.~S. Han, ``Hypothesis testing with multiterminal data compression,'' {\em
  IEEE Trans.~Inf.~Theory}, vol.~33, pp.~759--772, Nov. 1987.

\bibitem{SHA}
H.~Shimokawa, T.~Han, and S.~I. Amari, ``Error bound for hypothesis testing
  with data compression,'' in {\em Proc.~ISIT}, p.~114, Jul. 1994.

\bibitem{Wagner}
M.~S. Rahman and A.~B. Wagner, ``On the optimality of binning for distributed
  hypothesis testing,'' {\em IEEE Trans.~Inf.~Theory}, vol.~58, pp.~6282--6303,
  Oct. 2012.

\bibitem{Kochman-MAC}
E.~Haim and Y.~Kochman, ``Binary distributed hypothesis testing via
  korner-marton coding,'' in {\em Proc.~IEEE Info.~Theory Work.~(ITW)}, 2016.

\bibitem{WeinbergerKochman}
N.~Weinberger and Y.~Kochman, ``On the reliability function of distributed
  hypothesis testing under optimal detection,'' {\em IEEE Transactions on
  Information Theory}, vol.~65, no.~8, pp.~4940--4965, 2019.

\bibitem{KochmanWang}
Y.~Kochman and L.~Wang, ``Improved random-binning exponent for distributed
  hypothesis testing,'' {\em IEEE Transactions on Information Theory},
  pp.~1--1, 2025.

\bibitem{SadafMicheleLigong}
S.~{Salehkalaibar}, M.~{Wigger}, and L.~{Wang}, ``Hypothesis testing over the
  two-hop relay network,'' {\em IEEE Trans.~Inf.~Theory}, vol.~65,
  pp.~4411--4433, Jul. 2019.

\bibitem{salehkalaibar2020hypothesisv1}
S.~Salehkalaibar, M.~Wigger, and L.~Wang, ``Hypothesis testing in multi-hop
  networks.'' [Online]. Available: \url{https://arxiv.org/abs/1708.05198v1},
  2017.

\bibitem{Michele3}
S.~Salehkalaibar, M.~Wigger, and R.~Timo, ``On hypothesis testing against
  independence with multiple decision centers,'' {\em IEEE Trans. Comm.},
  vol.~66, pp.~2409--2420, Jan. 2018.

\bibitem{Mhanna}
M.~Mhanna and P.~Piantanida, ``On secure distributed hypothesis testing,'' in
  {\em 2015 IEEE International Symposium on Information Theory (ISIT)},
  pp.~1605--1609, 2015.

\bibitem{8125176}
J.~Liao, L.~Sankar, V.~Y.~F. Tan, and F.~du~Pin~Calmon, ``Hypothesis testing
  under mutual information privacy constraints in the high privacy regime,''
  {\em IEEE Transactions on Information Forensics and Security}, vol.~13,
  no.~4, pp.~1058--1071, 2018.

\bibitem{8664261}
S.~B. Amor, A.~Gilani, S.~Salehkalaibar, and V.~Y.~F. Tan, ``Distributed
  hypothesis testing with privacy constraints,'' in {\em 2018 International
  Symposium on Information Theory and Its Applications (ISITA)}, pp.~742--746,
  2018.

\bibitem{TACI_HT}
S.~Sreekumar and D.~G\"und\"uz, ``Testing against conditional independence
  under security constraints,'' in {\em 2018 IEEE International Symposium on
  Information Theory (ISIT)}, pp.~181--185, 2018.

\bibitem{Faour}
S.~Faour, M.~Hamad, M.~Sarkiss, and M.~Wigger, ``Testing against independence
  with an eavesdropper,'' in {\em 2023 IEEE Information Theory Workshop (ITW)},
  pp.~277--282, 2023.

\bibitem{CovertDHT}
A.~Bounhar, M.~Sarkiss, and M.~Wigger, ``Covert distributed detection over
  discrete memoryless channels,'' in {\em 2024 IEEE International Symposium on
  Information Theory (ISIT)}, pp.~172--177, 2024.

\bibitem{HWS20}
M.~Hamad, M.~Wigger, and M.~Sarkiss, ``Cooperative multi-sensor detection under
  variable-length coding,'' in {\em 2020 IEEE Information Theory Workshop
  (ITW)}, pp.~1--5, 2021.

\bibitem{JSAIT}
S.~Salehkalaibar and M.~Wigger, ``Distributed hypothesis testing with
  variable-length coding,'' {\em IEEE J. Sel. Areas Inf. Th.}, vol.~1, no.~3,
  pp.~681--694, 2020.

\bibitem{Han_Kobayashi}
T.~Han and K.~Kobayashi, ``Exponential-type error probabilities for
  multiterminal hypothesis testing,'' {\em IEEE Trans.~Inf.~Theory}, vol.~35,
  no.~1, pp.~2--14, 1989.

\bibitem{Shalaby}
H.~Shalaby and A.~Papamarcou, ``Multiterminal detection with zero-rate data
  compression,'' {\em IEEE Transactions on Information Theory}, vol.~38, no.~2,
  pp.~254--267, 1992.

\bibitem{PierreMichele}
P.~Escamilla, M.~Wigger, and A.~Zaidi, ``Distributed hypothesis testing:
  cooperation and concurrent detection,'' {\em IEEE Transactions on Information
  Theory}, vol.~66, no.~12, pp.~7550--7564, 2020.

\bibitem{Watanabe_DHT}
S.~Watanabe, ``Neyman-pearson test for zero-rate multiterminal hypothesis
  testing,'' {\em IEEE Transactions on Information Theory}, vol.~64, no.~7,
  pp.~4923--4939, 2018.

\bibitem{Sreekumar}
S.~Sreekumar, C.~Hirche, H.-C. Cheng, and M.~Berta, ``Distributed quantum
  hypothesis testing under zero-rate communication constraints,'' 2025.

\bibitem{Deniz_DHT}
S.~Sreekumar and D.~G\"und\"uz, ``Distributed hypothesis testing over discrete
  memoryless channels,'' {\em IEEE Trans.~Inf.~Theory}, vol.~66, no.~4,
  pp.~2044--2066, 2020.

\bibitem{Michele_noisy_and_MAC}
S.~Salehkalaibar and M.~Wigger, ``Distributed hypothesis testing based on
  unequal-error protection codes,'' {\em IEEE Trans.~Inf.~Theory}, vol.~66,
  pp.~4150--41820, Jul. 2020.

\bibitem{Sadaf_BC}
S.~Salehkalaibar and M.~Wigger, ``Distributed hypothesis testing over noisy
  broadcast channels,'' {\em Information}, vol.~12, no.~7, p.~268, 2021.

\bibitem{Csiszarbook}
I.~{Csisz\'ar} and J.~K\"orner, {\em Information theory: coding theorems for
  discrete memoryless systems}.
\newblock Cambridge University Press, 2011.

\bibitem{csiszar_book}
I.~Csisz{\'a}r and J.~K{\"o}rner, {\em Information theory: coding theorems for
  discrete memoryless systems}.
\newblock Cambridge University Press, 2011.

\bibitem{MartonBU}
K.~Marton, ``A simple proof of the blowing-up lemma,'' {\em IEEE
  Trans.~Inf.~Theory}, vol.~32, pp.~445--446, May 1986.

\end{thebibliography}
\end{document}